\documentclass[twocolumn, journal]{IEEEtran}
\IEEEoverridecommandlockouts

\usepackage{color}
\usepackage{xcolor}
\usepackage{bm}
\usepackage{graphicx}
\usepackage{amsmath}
\usepackage{amssymb}
\usepackage{algorithm}
\usepackage{algorithmic}
\usepackage{multirow}
\usepackage{booktabs}
\usepackage{array}
\usepackage{amsthm}
\usepackage{lipsum}
\usepackage{enumerate}
\usepackage{stfloats}
\usepackage{subfigure}
\usepackage{cases}
\usepackage{diagbox}
\usepackage{threeparttable}
\usepackage{cite}

\newcommand{\non}{\nonumber}

\allowdisplaybreaks[4]

\title{Channel Estimation and Beamforming for Beyond Diagonal Reconfigurable Intelligent Surfaces}

\author{Hongyu Li,~\IEEEmembership{Graduate Student Member,~IEEE}, Shanpu Shen,~\IEEEmembership{Senior Member,~IEEE}, \\Yumeng Zhang,~\IEEEmembership{Graduate Student Member,~IEEE}, and Bruno Clerckx,~\IEEEmembership{Fellow,~IEEE}
\thanks{Manuscript received; Part of this work has been accepted by IEEE International Workshop on Computational Advances in Multi-Sensor Adaptive Processing (CAMSAP), 2023 \cite{li2023channel}.}
\thanks{Hongyu Li, Yumeng Zhang, and Bruno Clerckx are with the Department of Electrical and Electronic Engineering, Imperial College London, London SW7 2AZ, U.K. (e-mail:\{c.li21,~yumeng.zhang19,~b.clerckx\}@imperial.ac.uk).}\\
\thanks{Shanpu Shen is with the Department of Electrical Engineering and Electronics, University of Liverpool, Liverpool L69 3GJ, U.K. (e-mail: Shanpu.Shen@liverpool.ac.uk).}
}

\begin{document}

\maketitle
\thispagestyle{empty}
\begin{abstract}
	Beyond diagonal reconfigurable intelligent surface (BD-RIS) is a new advance and generalization of the RIS technique. BD-RIS breaks through the isolation between RIS elements by creatively introducing inter-element connections, thereby enabling smarter wave manipulation and enlarging coverage. However, exploring proper channel estimation schemes suitable for BD-RIS aided communication systems still remains an open problem. 
    In this paper, we study channel estimation and beamforming design for BD-RIS aided multi-antenna systems. 
    We first describe the channel estimation strategy based on the least square (LS) method, derive the mean square error (MSE) of the LS estimation, and formulate the joint pilot sequence and BD-RIS design problem with unique constraints induced by BD-RIS architectures. 
    Specifically, we propose an efficient pilot sequence and BD-RIS design which theoretically guarantees to achieve the minimum MSE. 
    With the estimated channel, we then consider two BD-RIS scenarios and propose beamforming design algorithms. 
    Finally, we provide simulation results to verify the effectiveness of the proposed channel estimation scheme and beamforming design algorithms. 
    We also show that more inter-element connections in BD-RIS improves the performance while increasing the training overhead for channel estimation.
\end{abstract}


\section{Introduction}
\label{sec:intro}

Beyond diagonal reconfigurable intelligent surface (BD-RIS) is an emerging technique, which generalizes and goes beyond conventional RIS with diagonal phase shift matrix \cite{di2020smart,wu2021intelligent,pan2022overview} and generates scattering matrices not limited to being diagonal, by introducing connections among RIS elements \cite{li2023reconfigurable}.
Thanks to the flexible inter-element connections, BD-RIS has benefits in providing smarter wave manipulation and enlarging coverage compared to conventional RIS \cite{li2023reconfigurable}. 

Existing works of BD-RIS have been carried out for modeling \cite{shen2021}, beamforming design \cite{nerini2021reconfigurable,nerini2022optimal}, and mode/architecture design \cite{li2022,li2022beyond,nerini2023beyond}. 
The modeling of BD-RIS and the concept of group- and fully-connected architectures, named according to the circuit topology of inter-element connections, are first proposed in \cite{shen2021}, followed by the discrete-value design \cite{nerini2021reconfigurable} and optimal beamforming design \cite{nerini2022optimal}.
Inspired by the group/fully-connected architectures \cite{shen2021} and the concept of intelligent omni surface (IOS) with enlarged coverage \cite{zhang2022intelligent,zhang2022intelligent1}, BD-RIS with hybrid transmitting and reflecting mode \cite{li2022} and multi-sector mode \cite{li2022beyond} are proposed to achieve full-space coverage with enhanced channel gain and system performance.
To find a better performance-complexity trade-off of BD-RIS, BD-RIS with tree- and forest-connected architectures are proposed in \cite{nerini2023beyond}, which are proved to achieve the performance upper bound with the minimum circuit complexity.
The enhanced performance achieved by BD-RIS with different modes and architectures highly depends on accurate channel state information (CSI), while none of the above-mentioned works \cite{shen2021,li2022,li2022beyond,nerini2023beyond,nerini2021reconfigurable,nerini2022optimal} study the channel estimation of BD-RIS. 
Therefore, it remains an open problem to effectively obtain the CSI for BD-RIS aided wireless systems.

In conventional RIS literature on channel estimation, there are in general two strategies to obtain the instantaneous CSI \cite{zheng2022survey}.
The first strategy is to \textit{separately} estimate the channels between the base station (BS)/users and the RIS by partially ``activating'' RIS elements using RF chains. In this strategy, the separate channels can be estimated by leveraging channel reciprocity and time division duplex (TDD) and applying conventional estimation approaches, such as the estimation of signal parameters via rotational invariance technique (ESPRIT) and multiple signal classification (MUSIC) \cite{hu2021semi}, and compressed sensing (CS) \cite{alexandropoulos2020hardware,taha2019deep}. 
This strategy generally has low training overhead regardless of the number of RIS elements, and can be directly used in BD-RIS aided communication systems, since the structures of separate channels do not depend on RIS architectures.
Once the channel estimates are obtained, the existing BD-RIS design algorithms \cite{nerini2022optimal,li2022,li2022beyond} based on perfect and separate channels can still be utilized.  
However, the drawback is that the equipment of RF chains will result in additional cost and power consumption, which violates the original motivation of employing RIS in wireless communication systems. 
The second strategy is to estimate the \textit{cascaded} user-RIS-BS channels relying on purely passive RISs \cite{swindlehurst2022channel}.
In this strategy, the BS estimates the cascaded channel through proper design of the pilot sequence and RIS patterns, such as the ON/OFF-based design \cite{yang2020intelligent} and the orthogonality-based design \cite{zheng2019intelligent,you2020channel}.
To reduce the training overhead, other estimation schemes, such as a three-phase framework \cite{wang2020channel} and an anchor-assisted channel estimation \cite{guan2021anchor}, are further proposed making use of the common BS-RIS channels for different users. 
With the same motivation, a novel cascaded channel estimation scheme is proposed by exploiting the sparsity and correlations of millimeter wave channels \cite{zhou2022channel}. 
However, adapting this strategy to BD-RIS aided scenarios raises the following challenges:
\textit{First}, the structure of the cascaded channels is tightly coupled to the BD-RIS architectures and thus different from conventional RIS cases. As a result, the existing channel estimation schemes cannot be directly used for BD-RIS aided scenarios.
\textit{Second}, the dimension of the cascaded channel is larger than that for conventional RIS cases due to the inter-element connections of BD-RIS, such that significant training overhead for channel estimation is required.
\textit{Third}, the BD-RIS training pattern should also be re-designed due to the different constraints on the scattering matrix induced by different BD-RIS architectures.
\textit{Fourth}, existing BD-RIS design algorithms \cite{nerini2022optimal,li2022} based on separate channels are not applicable, such that algorithms suitable for cascaded channels should be developed.
Therefore, it remains a challenging problem to develop efficient channel estimation methods for BD-RIS aided scenarios with affordable overhead and to study corresponding beamforming design algorithms with cascaded channel estimates.

To address the above challenges, in this work, we study the channel estimation and beamforming design for BD-RIS aided wireless communication systems.
Specifically, we investigate the cascaded channel estimation for BD-RIS aided wireless systems. With only the cascaded channel estimate, we re-consider the BD-RIS design and propose corresponding algorithms for different scenarios. 
The contributions of this work are summarized as follows.

\textit{First}, we propose a novel channel estimation scheme for a reflective BD-RIS aided multiple input multiple output (MIMO) system, where the joint design of pilot sequence and BD-RIS matrix during the training process is investigated.
The proposed pilot sequence and BD-RIS design differs from that for conventional RIS cases due to the new structure of the cascaded channel and new constraints originating from the unique BD-RIS architectures.
The proposed design is applicable for general time-invariant channel models, such as commonly used i.i.d Rayleigh fading and Rician fading channels.
Specifically, we propose a closed-form pilot sequence and BD-RIS design which theoretically guarantees to achieve the minimum mean square error (MSE) of the least square (LS) estimator.
A comprehensive analysis of the training overhead of the proposed channel estimation scheme is also provided.

\textit{Second}, we generalize the proposed channel estimation scheme to different BD-RIS aided scenarios. On one hand, we show that the proposed channel estimation scheme can be used for reflective BD-RIS aided multi-user systems. On the other hand, we illustrate that the proposed scheme is also readily generalized to BD-RIS with hybrid transmitting and reflecting mode and multi-sector mode by assuming each sector of the hybrid/multi-sector BD-RIS works in turns.

\textit{Third}, we investigate the beamforming design for a reflective BD-RIS aided point-to-point MIMO system based on the cascaded channel estimates. The proposed design differs from \cite{nerini2022optimal} since only the CSI for cascaded channels is required.

\textit{Fourth}, we study the beamforming design for a hybrid and multi-sector BD-RIS aided multi-user multiple input single output (MU-MISO) system with cascaded channel estimates. The proposed design is a generalization of the algorithm in \cite{li2022}, with modifications adapting to the cascaded CSI.

\textit{Fifth}, we present simulation results to verify the effectiveness and accuracy of the proposed channel estimation scheme and evaluate the performance of the proposed beamforming design algorithms.
Simulation results verify that the proposed channel estimation scheme achieves the theoretical minimum MSE.
Simulation results also show that with the estimated channel obtained by the proposed channel estimation scheme, the proposed beamforming design algorithms can achieve satisfactory performance close to the perfect CSI cases. 
More importantly, there exists a practical trade-off between the channel estimation performance and the rate performance, indicating that the improved beamforming flexibility of BD-RIS architectures is achieved at the cost of increasing the training overhead.

This work has following contributions beyond \cite{li2023channel}: 1) The proposed channel estimation method is extended from the single receive antenna system to more general multi-antenna systems. 2) The proposed method is extended to multiuser scenarios and BD-RIS with different modes. 3) Beamforming design algorithms based on the channel estimates are proposed. 4) Analysis for the trade-off between channel estimation overhead and data transmission performance is provided.

\textit{Organization:} Section \ref{sec:syst_mod} describes the channel model and transmission protocol for a BD-RIS aided communication system. Section \ref{sec:CE} introduces the proposed channel estimation scheme for BD-RIS. Section \ref{sec:beamforming} illustrates the beamforming design of BD-RIS based on estimated channels. Section \ref{sec:simulation} evaluates the performance of the proposed channel estimation and beamforming design. Section \ref{sec:conclusion} concludes this work.

\textit{Notations:}
Boldface lower- and upper-case letters indicate vectors and matrices, respectively.
$\mathbb{C}$, $\mathbb{R}$, and $\mathbb{N}$ denote the sets of complex numbers, real numbers, and natural numbers, respectively. 
$\mathbb{E}\{\cdot\}$ represents the statistical expectation.
$\Re\{\cdot\}$ denotes the real part of complex numbers. 
$(\cdot)^T$, $(\cdot)^*$, $(\cdot)^H$, and $(\cdot)^{-1}$ denote the transpose, conjugate, conjugate-transpose, and inversion operations, respectively.
$\otimes$ and $\odot$ denote the Kronecker product and Hadamard product, respectively. 
$\mathsf{blkdiag}(\cdot)$ represents a block-diagonal matrix.
$|\cdot|$, $\|\cdot\|_2$, and $\|\cdot\|_F$ denote the absolute-value norm, the Euclidean norm, and the  Frobenius norm, respectively. 
$\mathsf{vec}(\cdot)$, $\mathsf{rank}(\cdot)$, and $\mathsf{tr}(\cdot)$, respectively, are the vectorization, rank, and trace of a matrix.
$\mathsf{unvec}(\cdot)$ is the reverse operation of the vectorization. Given a matrix $\mathbf{A}\in\mathbb{C}^{M\times N}$, we have $\mathbf{a}=\mathsf{vec}(\mathbf{A})\in\mathbb{C}^{MN\times 1}$ and $\mathbf{A} = \mathsf{unvec}(\mathbf{a})$.
$\mathsf{circshift}(\mathbf{a},N)$ rearranges vector $\mathbf{a}$ by moving the final $N$ entries to the first $N$ positions. 
$\mathsf{mod}(M,N)$ returns the remainder after $M$ is divided by $N$. 
$\mathbf{I}_M$ denotes an $M\times M$ identity matrix and $\mathbf{1}_{M}$ denotes an $M\times 1$ all-one vector.
$\mathbf{F}_M$ and $\mathbf{D}_M$, respectively, denote $M\times M$ discrete Fourier transform (DFT) matrix and Hadamard matrix. 
$[\mathbf{A}]_{i:i',j:j'}$ extracts the $i$-th to $i'$-th rows and the $j$-th to $j'$-th columns of $\mathbf{A}$. 
$[\mathbf{a}]_{i:i'}$ extracts the $i$-th to $i'$-th entries of $\mathbf{a}$.

\section{Channel Model and Transmission Protocol}
\label{sec:syst_mod}

In this section, we illustrate the uplink and downlink channel models of general BD-RIS assisted MIMO systems, and introduce the transmission protocol of the whole system. 

\subsection{Channel Model}

\begin{figure}
    \centering
    \includegraphics[width=0.48\textwidth]{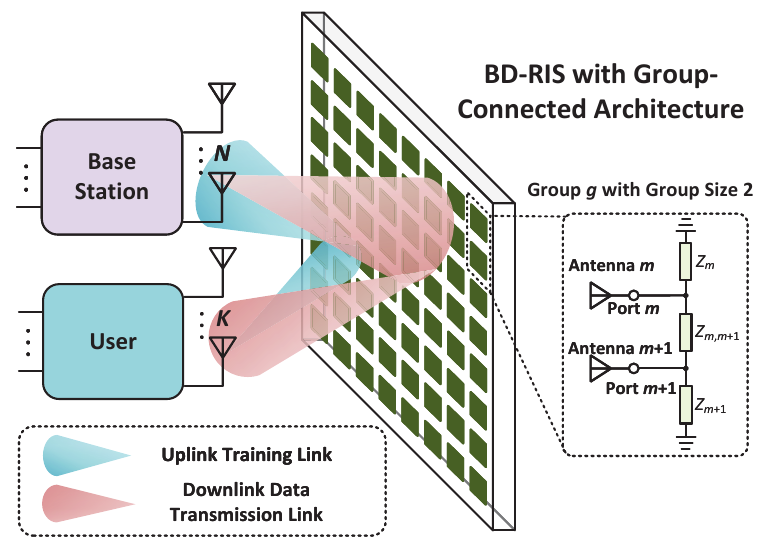}
    \caption{A paradigm of a BD-RIS assisted MIMO system.}
    \label{fig:syst_mod}
\end{figure}

We consider a narrowband system which consists of an $N$-antenna BS, an $M$-antenna BD-RIS with reflective mode, and a $K$-antenna user, as illustrated in Fig. \ref{fig:syst_mod}. 
The $M$ antennas of the BD-RIS are connected to an $M$-port group-connected reconfigurable impedance network \cite{shen2021}, where the $M$ ports are uniformly divided into $G_1$ groups with each containing $\bar{M} = \frac{M}{G_1}$ ports connected to each other\footnote{In practice, the uniform linear or planar array can be utilized for antenna deployment. Besides, varactors or PIN diodes can be used to implement the reconfigurable impedance network.}.
Mathematically, the BD-RIS with group-connected architecture has a block-diagonal scattering matrix\footnote{The group-connected architecture is a general case including both single- and fully-connected architectures as special cases. Specifically, $G_1=M$ boils down to single-connected architecture with diagonal scattering matrices; $G_1=1$ represents fully-connected architecture with full scattering matrices.} $\bar{\mathbf{\Phi}}\in\mathbb{C}^{M\times M}$ with each block $\bar{\mathbf{\Phi}}_g\in\mathbb{C}^{\bar{M}\times\bar{M}}$, $\forall g\in\mathcal{G}_1=\{1,\ldots,G_1\}$, being unitary \cite{shen2021}, i.e., 
\begin{equation}
    \bar{\mathbf{\Phi}} = \mathsf{blkdiag}(\bar{\mathbf{\Phi}}_1,\ldots,\bar{\mathbf{\Phi}}_{G_1}),  
\bar{\mathbf{\Phi}}_g^H\bar{\mathbf{\Phi}}_g = \mathbf{I}_{\bar{M}}, \forall g\in\mathcal{G}_1.
\label{eq:block_uni}
\end{equation}
In practice, the unitary matrix $\bar{\mathbf{\Phi}}_g$ might involve further constraints determined by the circuit topology of the reconfigurable impedance network, such that its entries are jointly controlled by specific numbers of the reconfigurable impedance components. More details about different circuit designs and corresponding mathematical constraints can be found in \cite{shen2021,nerini2023beyond}.

In this work, we assume the direct user-BS channel is blocked and focus purely on the estimation of the cascaded user-RIS-BS channel\footnote{When the direct BS-user channel exists, the direct channel can be effectively obtained by turning off the BD-RIS and using conventional channel estimation strategies.}.
In the uplink scenario, we denote $\mathbf{G}\in\mathbb{C}^{N\times M}$ and $\mathbf{H}\in\mathbb{C}^{M\times K}$ as the channel from the BD-RIS to BS, and from the user to BD-RIS, respectively. The uplink user-RIS-BS channel $\mathbf{H}_\mathrm{u}\in\mathbb{C}^{N\times K}$ is
\begin{equation}
    \label{eq:channel}
    \begin{aligned}
    \mathbf{H}_\mathrm{u} &= \mathbf{G}\bar{\mathbf{\Phi}}\mathbf{H} = \sum_{g\in\mathcal{G}_1}\mathbf{G}_g\bar{\mathbf{\Phi}}_g\mathbf{H}_g\\
    &= \mathsf{unvec}\Big(\sum_{g\in\mathcal{G}_1}\underbrace{\mathbf{H}_g^T\otimes\mathbf{G}_g}_{=\bar{\mathbf{Q}}_g\in\mathbb{C}^{NK\times\bar{M}^2}}\mathsf{vec}(\bar{\mathbf{\Phi}}_g)\Big),
    \end{aligned}
\end{equation}
where $\mathbf{G}_g = [\mathbf{G}]_{:,(g-1)\bar{M}+1:g\bar{M}}\in\mathbb{C}^{N\times\bar{M}}$ and $\mathbf{H}_g = [\mathbf{H}]_{(g-1)\bar{M}+1:g\bar{M},:}\in\mathbb{C}^{\bar{M}\times K}$, $\forall g\in\mathcal{G}_1$. 
Alternatively, in the downlink scenario, we denote $\mathbf{G}' = [\mathbf{G}_1'^T,\ldots,\mathbf{G}_{G_1}'^T]^T\in\mathbb{C}^{M\times N}$ and $\mathbf{H}' = [\mathbf{H}_1',\ldots,\mathbf{H}_{G_1}']\in\mathbb{C}^{K\times M}$, respectively, as the channels from BS to BD-RIS, and from BD-RIS to the user.
The downlink BS-RIS-user channel $\mathbf{H}_\mathrm{d}\in\mathbb{C}^{K\times N}$ is 
\begin{equation}
    \label{eq:downlink_channel}
    \begin{aligned}
        \mathbf{H}_\mathrm{d} &\overset{\text{(a)}}{=} \mathbf{H}'\bar{\mathbf{\Theta}}\mathbf{G}' = \sum_{g\in\mathcal{G}_1}\mathbf{H}_g'\bar{\mathbf{\Theta}}_g\mathbf{G}_g'\\
        &\overset{\text{(b)}}{=} \mathsf{unvec}^T\Big(\sum_{g\in\mathcal{G}_1}(\mathbf{H}_g'\otimes\mathbf{G}_g'^T)\mathsf{vec}(\bar{\mathbf{\Theta}}_g^T)\Big),
    \end{aligned}
\end{equation}
where (a) holds by defining the BD-RIS matrix for downlink data transmission as $\bar{\mathbf{\Theta}} = \mathsf{blkdiag}(\bar{\mathbf{\Theta}}_1,\ldots,\bar{\mathbf{\Theta}}_{G_1})$; (b) holds by the property of vectorization.

When the communication occurs with TDD and the reciprocity between the uplink and downlink user-RIS and RIS-BS channels exists, i.e., $\mathbf{H}'=\mathbf{H}^T$ and $\mathbf{G}'=\mathbf{G}^T$, we have 
\begin{equation}
    \mathbf{H}_\mathrm{d} = \mathsf{unvec}^T\Big(\sum_{g\in\mathcal{G}_1}\bar{\mathbf{Q}}_g\mathsf{vec}(\bar{\mathbf{\Theta}}_g^T)\Big),
\end{equation}
which indicates that the channel estimate $\bar{\mathbf{Q}} = [\bar{\mathbf{Q}}_1,\ldots,\bar{\mathbf{Q}}_{G_1}]\in\mathbb{C}^{NK\times\bar{M}^2G_1}$ obtained by the uplink training can be used for downlink data transmission\footnote{We should clarify here that $\bar{\mathbf{\Theta}} \ne \bar{\mathbf{\Phi}}^T$ since the BD-RIS matrices are designed focusing on different metrics, i.e., minimizing the channel estimation error and maximizing the data transmission rate.}. 
This motivates us to estimate $\bar{\mathbf{Q}}$ instead of separate channels $\mathbf{H}$ and $\mathbf{G}$, and design the BD-RIS beamforming with the knowledge of $\bar{\mathbf{Q}}$ for data transmission. 

\textit{Remark 1:} 
From (\ref{eq:channel}) and (\ref{eq:downlink_channel}) we observe that the BD-RIS aided cascaded channel for both uplink and downlink, i.e., $\bar{\mathbf{Q}}$, is constructed relying on the non-zero parts of the scattering matrix at BD-RIS, i.e., $\bar{\mathbf{\Phi}}$ or $\bar{\mathbf{\Theta}}$, which characterizes inter-element connections in BD-RIS architectures. That is, the channel to be estimated depends on the BD-RIS architectures. 
In addition, estimating $\bar{\mathbf{Q}}$ relies on the variation of $\bar{\mathbf{\Phi}}$ during the uplink training, with new constraints (\ref{eq:block_uni}) instead of being diagonal as in conventional RIS. 
As a result, the channel estimation schemes for conventional RIS \cite{yang2020intelligent,zheng2019intelligent,you2020channel} do not work for BD-RIS aided scenarios. Therefore, new channel estimation designs adaptive to BD-RIS architectures are needed.

\subsection{Tile-Based Channel Construction}

\begin{figure}
    \centering
    \includegraphics[width=0.48\textwidth]{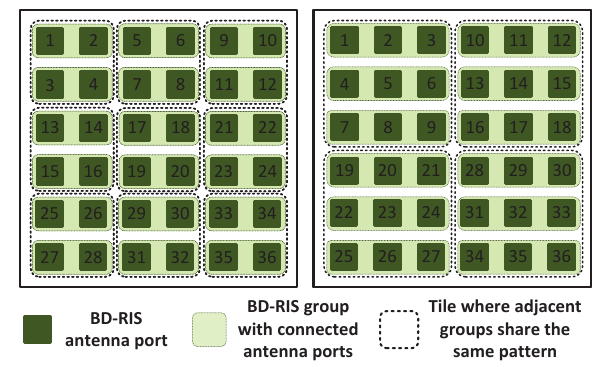}
    \caption{Illustration of BD-RIS antenna ports, groups, and tiles. Left: BD-RIS with $M=36$, $G_1=18$ and $G_2=9$; right: BD-RIS with $M=36$, $G_1=12$ and $G_2=4$. For example, in the left figure, BD-RIS antenna ports with indices $\{1,2\}$ and $\{3,4\}$ form two different groups, each of which consists of antenna ports connected to each other by reconfigurable impedance components. These two groups further form one tile, within which the groups share the same BD-RIS pattern.}
    \label{fig:ele_uni_group}
\end{figure}

Estimating $\bar{\mathbf{Q}}$ requires extremely high complexity and training overhead due to the large number of channel coefficients growing linearly with $G_1$ and quadratically with $\bar{M}$.  
To effectively reduce the channel estimation complexity and training overhead, we generalize the methods used in conventional RIS channel estimation works \cite{yang2020intelligent,kundu2022optimal,li2023performance} by introducing the concept of \textit{tile}, which consists of adjacent \textit{groups}, each having $\bar{M}$ ports connected to each other. 
The groups in the same tile share the same BD-RIS pattern during the training.
To ease understanding, we provide two examples in Fig. \ref{fig:ele_uni_group} to illustrate antenna ports, groups, and tiles.
Here, we define $G_2$ as the number of tiles and $\bar{G}=\frac{G_1}{G_2}$ as the tile size. 
We further define the shared pattern $\mathbf{\Phi}_i\in\mathbb{C}^{\bar{M}\times\bar{M}}$ for tile $i$ with $\mathbf{\Phi}_i^H\mathbf{\Phi}_i=\mathbf{I}_{\bar{M}}$, $\forall i\in\mathcal{G}_2=\{1,\ldots,G_2\}$ and rewrite the BD-RIS matrix $\bar{\mathbf{\Phi}}$ as 
\begin{equation}
    \bar{\mathbf{\Phi}} = \mathsf{blkdiag}(\mathbf{I}_{\bar{G}}\otimes\mathbf{\Phi}_1,\ldots,\mathbf{I}_{\bar{G}}\otimes\mathbf{\Phi}_{G_2}).
\end{equation}
The user-RIS-BS channel (\ref{eq:channel}) can thus be re-written as 
\begin{equation}
    \mathbf{H}_\mathrm{u} = \mathsf{unvec}\Big(\sum_{i\in\mathcal{G}_2} \underbrace{\sum\nolimits_{j=1}^{\bar{G}} \bar{\mathbf{Q}}_{(i-1)\bar{G}+j}}_{=\mathbf{Q}_i\in\mathbb{C}^{NK\times\bar{M}^2}}\mathsf{vec}(\mathbf{\Phi}_i)\Big).
\end{equation}
As such, we estimate the combined cascaded channel of each tile, referred to as the tile-based channel. Specifically, a larger tile size leads to the reduction of training overhead since the channel parameters to be estimated reduces from $\bar{\mathbf{Q}}$ to $\mathbf{Q} =[\mathbf{Q}_1,\ldots,\mathbf{Q}_{G_2}]\in\mathbb{C}^{NK\times\bar{M}^2G_2}$. 
However, this is achieved at the expense of obtaining reduced CSI, which will limit the beamforming flexibility of BD-RIS and further degrades the performance for data transmission.
In other words, although we focus on different objective functions for uplink channel estimation and downlink data transmission, the two designs are tightly related to each other due to the tile-based channel construction.
The impact of estimating the tile-based channel instead of the original cascaded channel on the beamforming flexibility of BD-RIS is mathematically reflected in the downlink channel $\mathbf{H}_\mathrm{d}$, which is re-written as 
\begin{equation}
    \label{eq:downlink_channel_stack}
    \mathbf{H}_\mathrm{d} = \mathsf{unvec}^T\Big(\sum_{i\in\mathcal{G}_2}\mathbf{Q}_i\mathsf{vec}(\mathbf{\Theta}_i^T)\Big),
\end{equation}
with a reformulation of the BD-RIS matrix as 
\begin{equation}
    \label{eq:theta_stack}
    \bar{\mathbf{\Theta}} = \mathsf{blkdiag}(\mathbf{I}_{\bar{G}}\otimes\mathbf{\Theta}_1,\ldots,\mathbf{I}_{\bar{G}}\otimes\mathbf{\Theta}_{G_2}).
\end{equation} 
This indicates that with the knowledge of $\mathbf{Q}$, we should design $\mathbf{\Theta}_i$, $i\in\mathcal{G}_2$, which provides less beam manipulation flexibility compared with designing $\bar{\mathbf{\Theta}}_g$, $\forall g\in\mathcal{G}_1$ as in perfect CSI cases, resulting in performance loss for downlink data transmission.

\textit{Remark 2:}
The trade-off between the training overhead for channel estimation and the performance for data transmission comes from twofold perspectives. 
On the one hand, due to the tile-based channel construction, the larger $G_2$ reduces the dimension of channels to be estimated, i.e., from $\bar{\mathbf{Q}}$ to $\mathbf{Q}$, which further reduces the overhead. 
Meanwhile, the larger $G_2$ reduces the dimension of BD-RIS matrices to be designed, i.e., from $\bar{\mathbf{\Theta}}_g$, $\forall g\in\mathcal{G}_1$ to $\mathbf{\Theta}_i$, $\forall i\in\mathcal{G}_2$. 
On the other hand, due to the BD-RIS architectures, the larger $\bar{M}$ increase the dimension of channels to be estimated and further the overhead. 
Meanwhile, the larger $\bar{M}$ provides more flexibility to BD-RIS design, which further enhances the performance.
This overhead-performance trade-off will be qualitatively and quantitatively studied in the following sections.

\subsection{Transmission Protocol}

\begin{figure}[t]
    \centering
    \includegraphics[width=0.48\textwidth]{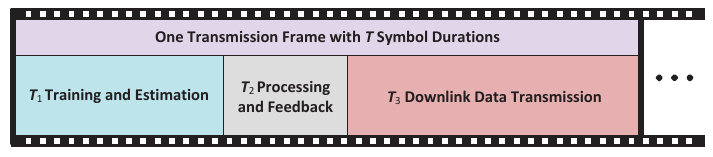}
    \caption{One transmission frame consisting of three phases with $T = T_1 + T_2 + T_3$ symbol durations.}
    \label{fig:protocol}
\end{figure}

With the tile-based channel construction, we instead estimate the reduced dimensional cascaded channel $\mathbf{Q}$ and design the beamforming with the estimate of $\mathbf{Q}$ for data transmission.
This results in the following protocol \cite{yang2020intelligent,swindlehurst2022channel,you2020channel} with each transmission frame divided into three phases as illustrated in Fig. \ref{fig:protocol}, based on the assumption that user-RIS and RIS-BS channels remain approximately constant within one transmission frame. 
In this protocol, we propose channel estimation and beamforming designs which are feasible for general time-invariant channel models, while the performance of the proposed designs are evaluated based on commonly-used Rician fading models in Section \ref{sec:simulation}.

\textit{Phase 1:} The BS estimates $\mathbf{Q}$ by uplink training, where the pilots are consecutively transmitted from the user and reflected by the BD-RIS with varied $\mathbf{\Phi}_i$, $\forall i\in\mathcal{G}_2$ within $T_1$ symbol durations.  
In this phase, the difference compared to conventional RIS comes from the design of the varied BD-RIS pattern due to the different constraint of the scattering matrix, whose details will be given in Section \ref{sec:CE}.

\textit{Phase 2:} The BS optimizes the transceiver beamformers, and $\mathbf{\Theta}_i$, $\forall i\in\mathcal{G}_2$ based on the estimated channel $\mathbf{Q}$, and feeds back the optimized $\mathbf{\Theta}_i$, $\forall i\in\mathcal{G}_2$ to BD-RIS requiring $T_2$ symbol durations. This phase is also different from conventional RIS cases due to the more general unitary constraint of the BD-RIS, whose details will be given in Section \ref{sec:beamforming}.

\textit{Phase 3:} The BS performs the downlink data transmission for $T_3$ symbol durations based on the optimized beamformers and $\mathbf{\Theta}_i$, $\forall i\in\mathcal{G}_2$ in Phase 2. 

\section{Channel Estimation for BD-RIS}
\label{sec:CE}

In this section, we describe the proposed channel estimation strategy based on the LS method, formulate the BD-RIS design problem to minimize the MSE of the LS estimator, and provide the closed-form solution.

\subsection{LS Based Channel Estimation}

The channel estimation strategy is described as follows. Assuming the user sends pilot symbol vector $\bar{\mathbf{x}}_{t}\in\mathbb{C}^{K\times 1}$ with $|[\bar{\mathbf{x}}_t]_k| = 1$, $\forall k\in\mathcal{K} = \{1,\ldots,K\}$, at time slot $t$, $\forall t\in\mathcal{T} = \{1,\ldots,T_1\}$, the signal received at the BS is 
\begin{equation}\label{eq:received_signal_uplink}
    \begin{aligned}
        \mathbf{y}_t &= \sqrt{P_\mathrm{u}}\mathbf{G}\bar{\bar{\mathbf{\Phi}}}_t\mathbf{H}\bar{\mathbf{x}}_t + \mathbf{n}_t\\
        &= \sqrt{P_\mathrm{u}}(\bar{\mathbf{x}}_t^T\otimes\mathbf{I}_N)\sum_{i\in\mathcal{G}_2}\mathbf{Q}_i\mathsf{vec}(\mathbf{\Phi}_{t,i}) + \mathbf{n}_t\\
        &= \sqrt{P_\mathrm{u}}(\bar{\mathbf{x}}_t^T\otimes\mathbf{I}_N)\mathbf{Q}\bar{\bm{\phi}}_t + \mathbf{n}_t\\
        &= \sqrt{P_\mathrm{u}}\underbrace{(\bar{\bm{\phi}}_t^T\otimes\bar{\mathbf{x}}_t^T\otimes\mathbf{I}_N)}_{=\widehat{\mathbf{\Phi}}_t\in\mathbb{C}^{N\times NK\bar{M}^2G_2}}\mathsf{vec}(\mathbf{Q}) + \mathbf{n}_t, \forall t\in\mathcal{T},
    \end{aligned}
\end{equation}
where $P_\mathrm{u}$ denotes the transmit power at each user antenna, $\bar{\bar{\mathbf{\Phi}}}_t = \mathsf{blkdiag}(\mathbf{I}_{\bar{G}}\otimes\mathbf{\Phi}_{t,1},\ldots,\mathbf{I}_{\bar{G}}\otimes\mathbf{\Phi}_{t,G_2})$ with $\mathbf{\Phi}_{t,i}\in\mathbb{C}^{\bar{M}\times\bar{M}}$ denoting the $i$-th tile of the BD-RIS matrix at time slot $t$, and $\mathbf{n}_t \in\mathbb{C}^{N\times 1}$ with $\mathbf{n}_t\sim\mathcal{CN}(\mathbf{0},\sigma^2\mathbf{I}_N)$, $\forall t\in\mathcal{T}$ denotes the noise.
The vector $\bar{\bm{\phi}}_t$ is defined as $\bar{\bm{\phi}}_t = [\mathsf{vec}^T(\mathbf{\Phi}_{t,1}),\ldots,\mathsf{vec}^T(\mathbf{\Phi}_{t,G_2})]^T\in\mathbb{C}^{G_2\bar{M}^2}$, $\forall t\in\mathcal{T}$.
To uniquely estimate the cascaded channel $\mathbf{Q}$ in (\ref{eq:received_signal_uplink}), $T_1\ge K\bar{M}^2G_2$ pilot symbols should be transmitted from the user. Combining the data from such pilots together, we have 
\begin{equation}
    \begin{aligned}
        \mathbf{y} &= [\mathbf{y}_1^T,\ldots,\mathbf{y}_{T_\mathrm{1}}^T]^T\\
        &= \sqrt{P_\mathrm{u}}\underbrace{[\widehat{\mathbf{\Phi}}_1^T,\ldots,\widehat{\mathbf{\Phi}}^T_{T_\mathrm{1}}]^T}_{=\widehat{\mathbf{\Phi}}\in\mathbb{C}^{NT_1\times NK\bar{M}^2G_2}}\mathbf{q} + \underbrace{[\mathbf{n}_1^T,\ldots,\mathbf{n}_{T_\mathrm{1}}^T]^T}_{=\mathbf{n}\in\mathbb{C}^{NT_1\times 1}},
    \end{aligned}
\end{equation}
where $\mathbf{q} = \mathsf{vec}(\mathbf{Q}) \in\mathbb{C}^{NK\bar{M}^2G_2}$.
The simplest way to estimate $\mathbf{q}$ is to use the LS method, yielding the LS estimator of $\mathbf{q}$ as the following form 
\begin{equation}
    \label{eq:channel_est}
    \begin{aligned}
        \widehat{\mathbf{q}} &= (\sqrt{P_\mathrm{u}})^{-1}\widehat{\mathbf{\Phi}}^\dagger\mathbf{y}
         = (\sqrt{P_\mathrm{u}})^{-1}(\widehat{\mathbf{\Phi}}^H\widehat{\mathbf{\Phi}})^{-1}\widehat{\mathbf{\Phi}}^H\mathbf{y}\\
          &= \mathbf{q} + \underbrace{(\sqrt{P}_\mathrm{u})^{-1}(\widehat{\mathbf{\Phi}}^H\widehat{\mathbf{\Phi}})^{-1}\widehat{\mathbf{\Phi}}^H\mathbf{n}}_{=\mathbf{q}_\mathrm{e}\in\mathbb{C}^{NK\bar{M}^2G_2\times 1}}.
    \end{aligned}
\end{equation} 
It is worth noting that the LS estimator treats each channel element as an independent unknown parameter, regardless of the strength of each channel coefficients and the correlations between coefficients. As a result, the channel estimation error does not depend on the channel models, but only on the transmit power $P_\mathrm{u}$, the concatenated pilot matrix $\widehat{\mathbf{\Phi}}$, and the noise.
The MSE of the LS estimator is 
\begin{equation}
    \label{eq:mse_ls_estimate}
    \mathrm{e}_{\widehat{\mathbf{q}}} = \mathbb{E}\{\|\widehat{\mathbf{q}} - \mathbf{q}\|_2^2\} = \mathbb{E}\{\|\mathbf{q}_\mathrm{e}\|_2^2\}=\frac{\sigma^2}{P_\mathrm{u}}\mathsf{tr}((\widehat{\mathbf{\Phi}}^H\widehat{\mathbf{\Phi}})^{-1}),
\end{equation} 
which implies that the MSE of channel estimation depends on the values of $\bar{\mathbf{x}}_t$ and $\bar{\bm{\phi}}_t$, $\forall t\in\mathcal{T}$, such that a proper design of them is required. 
As such, we formulate the following MSE minimization problem 
\begin{subequations}\label{eq:prob_MSE}
    \begin{align}
        \label{eq:obj_MSE}
        \min_{\{\bar{\mathbf{x}}_t,\bar{\bm{\phi}}_t\}_{\forall t}}~ &\mathsf{tr}((\widehat{\mathbf{\Phi}}^H\widehat{\mathbf{\Phi}})^{-1})\\
        \label{eq:unitary_constraint}
        \text{s.t.}~~~~ &\mathbf{\Phi}_{t,i}^H\mathbf{\Phi}_{t,i} = \mathbf{I}_{\bar{M}}, \forall t\in\mathcal{T},\forall i\in\mathcal{G}_2,\\
        \label{eq:pilot_constraint}
        &|[\bar{\mathbf{x}}_t]_k| = 1, \forall t\in\mathcal{T}, \forall k\in\mathcal{K},\\
        \label{eq:rank}
        &\mathsf{rank}(\widehat{\mathbf{\Phi}}) = NK\bar{M}^2G_2,
    \end{align}
\end{subequations}
where we set $T_1=K\bar{M}^2G_2$ to minimize the overhead without introducing estimation ambiguities, which yields $\widehat{\mathbf{\Phi}}$ a full-rank square matrix. 
Problem (\ref{eq:prob_MSE}) is difficult to solve due to the inverse operation in the objective and the non-convex constraints of group-connected BD-RIS. In the following subsection, we will first simplify the objective function and then propose an efficient approach for optimal $\bar{\mathbf{x}}_t$ and $\bar{\bm{\phi}}_t$, $\forall t\in\mathcal{T}$.

\subsection{Solution to Problem (\ref{eq:prob_MSE})}
We start by simplifying the inverse operation in objective (\ref{eq:obj_MSE}). To this end, we derive the minimum of the objective (\ref{eq:obj_MSE}) based on the following lemma. 

\textit{Lemma 1:}
The objective function in (\ref{eq:obj_MSE}) achieves its minimum, i.e., $\min \mathsf{tr}((\widehat{\mathbf{\Phi}}^H\widehat{\mathbf{\Phi}})^{-1}) = N\bar{M}$, with the condition $\widehat{\mathbf{\Phi}}^H\widehat{\mathbf{\Phi}}= \widehat{\mathbf{\Phi}}\widehat{\mathbf{\Phi}}^H= K\bar{M}G_2\mathbf{I}_{NK\bar{M}^2G_2}$.

\textit{Proof:}
Please refer to Appendix A. 
\hfill  $\square$

With Lemma 1, we deduce that achieving the minimum MSE in problem (\ref{eq:prob_MSE}) is equivalent to finding a scaled unitary matrix $\widehat{\mathbf{\Phi}}$, i.e., $\widehat{\mathbf{\Phi}}^H\widehat{\mathbf{\Phi}} = \widehat{\mathbf{\Phi}}\widehat{\mathbf{\Phi}}^H= K\bar{M}G_2\mathbf{I}_{NK\bar{M}^2G_2}$. Problem (\ref{eq:prob_MSE}) is thus transformed into a feasibility-check problem 
\begin{subequations}\label{eq:feasibility_check}
    \begin{align}
        \text{find} ~~&\widehat{\mathbf{\Phi}}\in\mathbb{C}^{NK\bar{M}^2G_2\times NK\bar{M}^2G_2}\\
        \label{eq:fc_cons1}
        \text{s.t.}~~ &\widehat{\mathbf{\Phi}}^H\widehat{\mathbf{\Phi}} = K\bar{M}G_2\mathbf{I}_{NK\bar{M}^2G_2},
        \text{(\ref{eq:unitary_constraint}), (\ref{eq:pilot_constraint})}.
    \end{align}
\end{subequations}

\begin{figure}[t]
    \centering
    \includegraphics[width=0.48\textwidth]{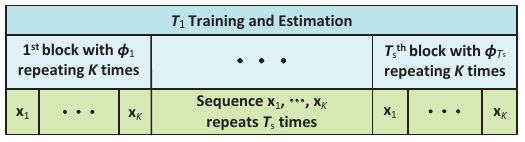}
    \caption{Training breaking the training time into $T_\mathrm{s}$ blocks of length $K$.}
    \label{fig:block_training}
\end{figure}

Given the large dimension and complicated structure of matrix $\widehat{\mathbf{\Phi}}$, which is constructed by a coupling of $\bar{\bm{\phi}}_t$ and $\bar{\mathbf{x}}_t$, $\forall t\in\mathcal{T}$, it is difficult to directly find such a matrix to simultaneously satisfy the three constraints in problem (\ref{eq:feasibility_check}). 
To decouple the design of $\bar{\bm{\phi}}_t$ and $\bar{\mathbf{x}}_t$, $\forall t\in\mathcal{T}$, we apply the most common training approach \cite{de2021channel}, which breaks the training time into $T_\mathrm{s} = \frac{T_1}{K} = \bar{M}^2G_2$ blocks of length $K$, as illustrated in Fig. \ref{fig:block_training}. 
Specifically, the same BD-RIS pattern, $\bm{\phi}_s$, is repeated $K$ times within each block $s\in\mathcal{S}=\{1,\ldots,T_\mathrm{s}\}$, while it varies from block to block. 
Meanwhile, the pilots $\mathbf{X} = [\mathbf{x}_1,\ldots,\mathbf{x}_K]^T$ are chosen as an orthogonal sequence, i.e., $\mathbf{X}^H\mathbf{X} = K\mathbf{I}_K$, and are repeated over the $T_\mathrm{s}$ blocks. 
In other words, we establish the BD-RIS pattern set $\bm{\phi}_s$, $\forall s\in\mathcal{S}$, and the pilot sequence set $\mathbf{x}_k$, $\forall k\in\mathcal{K}$, for periodicity.
Specifically, we have the mappings between $\bar{\bm{\phi}}_t$ and $\bm{\phi}_s$, and between $\bar{\mathbf{x}}_t$ and $\mathbf{x}_k$, respectively, as
\begin{equation}
    \bar{\bm{\phi}}_{(s-1)K+k} = \bm{\phi}_s, \bar{\mathbf{x}}_{(s-1)K+k} = \mathbf{x}_k, \forall k\in\mathcal{K}, \forall s\in\mathcal{S}. \label{eq:block_training}
\end{equation}
This allows us to rewrite $\widehat{\mathbf{\Phi}}^H\widehat{\mathbf{\Phi}}$ in the following form 
\begin{equation}
    \label{eq:simp_obj}
    \begin{aligned}
        \widehat{\mathbf{\Phi}}^H\widehat{\mathbf{\Phi}} &= \sum_{t=1}^{K\bar{M}^2G_2}(\bar{\bm{\phi}}_t^*\bar{\bm{\phi}}_t^T)\otimes(\bar{\mathbf{x}}_t^*\bar{\mathbf{x}}_t^T)\otimes\mathbf{I}_N\\
        &\overset{\text{(a)}}{=} \Big(\sum_{s=1}^{T_\mathrm{s}}\bm{\phi}_s^*\bm{\phi}_s^T\Big)\otimes\Big(\sum_{k=1}^K\mathbf{x}_k^*\mathbf{x}_k^T\Big)\otimes\mathbf{I}_N\\
        &\overset{\text{(b)}}{=} K(\mathbf{\Phi}^H\mathbf{\Phi})\otimes\mathbf{I}_{NK},
    \end{aligned}
\end{equation}
where (a) holds due to the mappings in (\ref{eq:block_training}); (b) holds with $\mathbf{X}^H\mathbf{X} = K\mathbf{I}_K$, and by defining $\mathbf{\Phi} = [\bm{\phi}_1,\ldots,\bm{\phi}_{T_\mathrm{s}}]^T\in\mathbb{C}^{\bar{M}^2G_2\times \bar{M}^2G_2}$. 
Based on (\ref{eq:simp_obj}), we transform designing $\bar{\bm{\phi}}_t$ and $\bar{\mathbf{x}}_t$ in (\ref{eq:feasibility_check}) into designing matrices $\mathbf{\Phi}$ and $\mathbf{X}$. Specifically, the feasible solution of $\mathbf{X}$ can be easily obtained, such as using DFT or Hadamard matrices, such that the $\bar{\mathbf{x}}_t$ is given by (\ref{eq:block_training}). The design of $\mathbf{\Phi}$, meanwhile, can be formulated as the following feasibility-check problem 
\begin{subequations}\label{eq:feasibility_check1}
    \begin{align}
        \text{find} ~~&\mathbf{\Phi}\in\mathbb{C}^{\bar{M}^2G_2\times \bar{M}^2G_2}\\
        \text{s.t.}~~ &\mathbf{\Phi}^H\mathbf{\Phi} = \bar{M}G_2\mathbf{I}_{\bar{M}^2G_2},
        \text{(\ref{eq:unitary_constraint})}.
    \end{align}
\end{subequations}

\textit{Remark 3:}
Problem (\ref{eq:feasibility_check1}) is a generalization of that for conventional RIS when $\bar{M} = 1$, in which case the solution can be directly obtained by DFT or Hadamard matrices. 
However, problem (\ref{eq:feasibility_check1}) is also a new formulation different from conventional RIS case and difficult to solve due to the constraint from BD-RIS architectures, resulting in the complicated construction of $\mathbf{\Phi}$ involving both inter- and intra-row orthogonality and having a large dimension.

To simplify the design of the large-dimensional matrix $\mathbf{\Phi}$, we propose to decouple problem (\ref{eq:feasibility_check1}) into two reduced-dimensional sub-problems with simpler constraints. This leads to the following lemma.

\textit{Lemma 2:} 
The feasible matrix $\mathbf{\Phi}$ from problem (\ref{eq:feasibility_check1}) can be constructed as $\mathbf{\Phi} = \mathbf{A}\otimes\breve{\mathbf{\Phi}}$, where $\mathbf{A}\in\mathbb{C}^{G_2\times G_2}$ is obtained by solving the following problem:
\begin{subequations}\label{eq:find_X}
    \begin{align}       
        \text{find} ~~&\mathbf{A}\in\mathbb{C}^{G_2\times G_2}\\
        \label{eq:fx_cons1}
        \text{s.t.} ~~&\mathbf{A}^H\mathbf{A} = G_2\mathbf{I}_{G_2},\\
        \label{eq:fx_cons2}
        &|[\mathbf{A}]_{i,i'}| = 1, \forall i,i'\in\mathcal{G}_2.
    \end{align}
\end{subequations} 
$\breve{\mathbf{\Phi}}\in\mathbb{C}^{\bar{M}^2\times\bar{M}^2}$ is obtained by solving the following problem:
\begin{subequations}\label{eq:find_Phy}
    \begin{align}       
        \text{find} ~~&\breve{\mathbf{\Phi}}\in\mathbb{C}^{\bar{M}^2\times\bar{M}^2}\\
        \label{eq:fc1_cons1}
        \text{s.t.} ~~&\breve{\mathbf{\Phi}}^H\breve{\mathbf{\Phi}} = \breve{\mathbf{\Phi}}\breve{\mathbf{\Phi}}^H = \bar{M}\mathbf{I}_{\bar{M}^2},\\
        \label{eq:fc1_cons2}
        &\breve{\mathbf{\Phi}}_m^H\breve{\mathbf{\Phi}}_m= \mathbf{I}_{\bar{M}}, \forall m\in\bar{\bar{\mathcal{M}}},
    \end{align}
\end{subequations}
where $\breve{\mathbf{\Phi}}_m = \mathsf{unvec}([\breve{\mathbf{\Phi}}^T]_{:,m})$ and $\bar{\bar{\mathcal{M}}} = \{1,\ldots,\bar{M}^2\}$.

\textit{Proof:} 
Please refer to Appendix B.
\hfill $\square$

With Lemma 2, we decouple problem (\ref{eq:feasibility_check1}) into two reduced-dimensional problems (\ref{eq:find_X}) and (\ref{eq:find_Phy}). 
The feasible solution to problem (\ref{eq:find_X}) can be easily obtained by using a $G_2\times G_2$ DFT matrix $\mathbf{F}_{G_2}$ or a Hadamard matrix $\mathbf{D}_{G_2}$, i.e., $\mathbf{A}^\mathrm{DFT} = \mathbf{F}_{G_2}$ or $\mathbf{A}^\mathrm{Hadamard} = \mathbf{D}_{G_2}$. However, the solution to problem (\ref{eq:find_Phy}) is not a straightforward extension from conventional RIS since we need to find a matrix $\breve{\mathbf{\Phi}}$ such that each row constructs a unitary matrix, i.e., constraint (\ref{eq:fc1_cons2}), and that different rows are orthogonal to each other, i.e., constraint (\ref{eq:fc1_cons1}). 
In other words, the matrix $\breve{\mathbf{\Phi}}$ should include two-dimensional (intra- and inter-row) orthogonality, which motivates us to construct $\breve{\mathbf{\Phi}}$ with two orthogonal bases, $\mathbf{Z}_1\in\mathbb{C}^{\bar{M}\times\bar{M}}$ and $\mathbf{Z}_2\in\mathbb{C}^{\bar{M}\times\bar{M}}$. The former guarantees the orthogonality within each row of $\breve{\mathbf{\Phi}}$; $\mathbf{Z}_2$ and the circular shift of $\mathbf{Z}_1$ guarantee that $\breve{\mathbf{\Phi}}$ has orthogonal rows.
This brings the following theorem. 

\textit{Theorem 1:}
The matrix $\breve{\mathbf{\Phi}}$ satisfying (\ref{eq:fc1_cons1}) and (\ref{eq:fc1_cons2}) can be constructed such that each row, i.e., $[\breve{\mathbf{\Phi}}]_{(m-1)\bar{M}+n,:} = \breve{\bm{\phi}}_{m,n}$, $\forall m,n\in\bar{\mathcal{M}} = \{1,\ldots,\bar{M}\}$, has the following structure:
\begin{equation}
    \label{eq:phi_construction}
    \begin{aligned}
    \breve{\bm{\phi}}_{m,n} = \mathsf{circshift}(\mathsf{vec}^T(\mathbf{Z}_1),(n-1)\bar{M})\odot([\mathbf{Z}_2]_{m,:}\otimes\mathbf{1}_{\bar{M}}^T),
    \end{aligned}
\end{equation}
where $\mathbf{Z}_1\in\mathbb{C}^{\bar{M}\times\bar{M}}$ and $\mathbf{Z}_2\in\mathbb{C}^{\bar{M}\times\bar{M}}$ are referred to as base matrices satisfying the following constraints: 
\begin{enumerate}[1)]
    \item $\mathbf{Z}_1$ is a scaled unitary matrix, i.e., $\mathbf{Z}_1^H\mathbf{Z}_1=\mathbf{Z}_1\mathbf{Z}_1^H=\alpha_1\mathbf{I}_{\bar{M}}$ with $\alpha\in\mathbb{R}$, $\alpha_1\ne 0$ a scalar; 
    \item $\mathbf{Z}_2$ is a scaled unitary matrix whose entries have identical modulus, i.e., $\mathbf{Z}_2^H\mathbf{Z}_2=\mathbf{Z}_2\mathbf{Z}_2^H=\alpha_2\mathbf{I}_{\bar{M}}$ with $\alpha_2\in\mathbb{R}$, $\alpha_2\ne 0$ a scalar, $|[\mathbf{Z}_2]_{m,n}|=\sqrt{\frac{\alpha_2}{\bar{M}}}$, $\forall m,n\in\bar{\mathcal{M}}$;
    \item The product of two scalars is a constant, i.e., $\alpha_1\alpha_2 = \bar{M}$.
\end{enumerate}

\textit{Proof:}
Please refer to Appendix C.
\hfill $\square$

Based on Theorem 1, we transform problem (\ref{eq:find_Phy}) into finding two base matrices $\mathbf{Z}_1$ and $\mathbf{Z}_2$ satisfying the above three constraints in Theorem 1, which 
can be easily satisfied using DFT/Hadamard matrices. 
More specifically, we can simply set $\mathbf{Z}_1^\mathrm{DFT} = \mathbf{F}_{\bar{M}}$ or $\mathbf{Z}_1^\mathrm{Hadamard} = \mathbf{D}_{\bar{M}}$ and $\mathbf{Z}_2^\mathrm{DFT} = \frac{1}{\sqrt{\bar{M}}}\mathbf{F}_{\bar{M}}$ or $\mathbf{Z}_2^\mathrm{Hadamard} = \frac{1}{\sqrt{\bar{M}}}\mathbf{D}_{\bar{M}}$ to construct $\breve{\mathbf{\Phi}}$, and further construct $\mathbf{\Phi} = \mathbf{A}\otimes\breve{\mathbf{\Phi}}$ with $\mathbf{A}^\mathrm{DFT}$ or $\mathbf{A}^\mathrm{Hadamard}$. 

With solutions to construct pilot sequence $\mathbf{X}$ and matrix $\mathbf{\Phi}$, the procedure of the proposed channel estimation scheme is straightforward, which is summarized as the following steps:
\begin{enumerate}[S1:]
    \item Generate pilot sequence $\mathbf{X}$ such that $\mathbf{X}^H\mathbf{X} = K\mathbf{I}_K$.
    \item Generate $\mathbf{A}$ by $\mathbf{A}^\mathrm{DFT} = \mathbf{F}_{G_2}$ or $\mathbf{A}^\mathrm{Hadamard}= \mathbf{D}_{G_2}$.
    \item Set $\mathbf{Z}_1$ by $\mathbf{Z}_1^\mathrm{DFT} = \mathbf{F}_{\bar{M}}$ or $\mathbf{Z}_1^\mathrm{Hadamard}= \mathbf{D}_{\bar{M}}$.
    \item Set $\mathbf{Z}_2$ by $\mathbf{Z}_2^\mathrm{DFT} = \frac{1}{\sqrt{\bar{M}}}\mathbf{F}_{\bar{M}}$ or $\mathbf{Z}_2^\mathrm{Hadamard}= \frac{1}{\sqrt{\bar{M}}}\mathbf{D}_{\bar{M}}$.
    \item Construct $\breve{\mathbf{\Phi}}$ by (\ref{eq:phi_construction}) from Theorem 1.
    \item Construct $\mathbf{\Phi}$ by $\mathbf{\Phi} = \mathbf{A}\otimes\breve{\mathbf{\Phi}}$ from Lemma 2.
    \item Map $\mathbf{X}$ and $\mathbf{\Phi}$ to $\bar{\mathbf{x}}_t$ and $\bar{\bm{\phi}}_t$, $\forall t\in\mathcal{T}$ by (\ref{eq:block_training}).
    \item Construct $\widehat{\mathbf{\Phi}}$ by (\ref{eq:received_signal_uplink}) with $\bar{\bm{\phi}}_t$ and $\bar{\mathbf{x}}_t$, $\forall t\in\mathcal{T}$.
    \item Obtain $\widehat{\mathbf{q}}$ by (\ref{eq:channel_est}).
\end{enumerate} 

\textit{Remark 4:}
Note that either using DFT matrices or Hadamard matrices to generate $\mathbf{\Phi}$, we achieve the minimum MSE of the LS estimation, i.e., 
$\mathrm{e}_{\widehat{\mathbf{q}}}^\mathrm{min} = N\sigma^2P_\mathrm{u}^{-1}\bar{M}$,
and avoid the time-consuming inversion by $\widehat{\mathbf{\Phi}}^\dag = (K\bar{M}G_2)^{-1}\widehat{\mathbf{\Phi}}^H$. Nevertheless, the two ways have pros and cons from different aspects. 
On one hand, the matrix $\mathbf{\Phi}$ constructed by Hadamard bases has the additional advantage of requiring only two states of each entry, which indicates that practical group-connected BD-RIS with discrete values can be effectively tuned to perform accurate channel estimation.
However, the values of $G_2$ and $\bar{M}$ should be chosen from the set $\{1,2,4n, n\in\mathbb{N}\}$ to guarantee the existance of Hadamard matrix.
On the other hand, $\mathbf{\Phi}$ constructed by DFT bases requires more states of each entry of group-connected BD-RIS, which complicates the practical implementation of BD-RIS, while there is no such limitation of the values of $G_2$ and $\bar{M}$.

\begin{table}[t]
    \renewcommand{\arraystretch}{1.25}
    \caption{Training Overhead, Estimation MSE, and Circuit Complexity of BD-RIS with Group-Connected Architectures}
    \centering 
    \begin{tabular}{|c|c|c|}
    \hline 
    Metrics & Value & Property\tabularnewline
    \hline 
    \hline
    Training Overhead & $T_1 = KM\bar{M}\bar{G}^{-1}$ & $\uparrow$ with $\bar{M}$; $\downarrow$ with $\bar{G}$ \\
    \hline
    Estimation MSE & $\mathrm{e}^\mathrm{min}_{\widehat{\mathbf{q}}} = \frac{\sigma^2}{P_\mathrm{u}}N\bar{M}$ & $\uparrow$ with $\bar{M}$\\
    \hline 
    Circuit Complexity & $C = (\bar{M}+1)\frac{M}{2}$ \cite{shen2021} & $\uparrow$ with $\bar{M}$ \\
    \hline 
    \end{tabular}
    \label{tab:trade_off}
\end{table}

\subsection{Overhead, MSE, and Circuit Complexity} 
The training overhead $T_1$ to estimate $\mathbf{q}$, the estimation error $\mathrm{e}^\mathrm{min}_{\widehat{\mathbf{q}}}$, and the circuit complexity $C$ of the group-connected BD-RIS indicated by the number of impedances to realize the circuit topology, are summarized in Table \ref{tab:trade_off}. 
We can observe that with fixed number of RIS antennas $M$ and other parameters, the training overhead, estimation MSE, and circuit complexity all increase with the group size $\bar{M}$ of the group-connected architecture. 
Interestingly, the training overhead can be effectively reduced by introducing tiles (increasing tile size $\bar{G}$), although the larger $\bar{M}$ decreases the number of possible $\bar{G}$.
More importantly, the larger $\bar{M}$ provides more beam manipulation flexibility to BD-RIS to enhance the transmission performance, while the larger $\bar{G}$ leads to reduced CSI and further limits the transmission performance. 
This phenomenon which will be numerically studied in Section \ref{sec:simulation}.
Therefore, it is important to properly choose the values of $\bar{M}$ and $\bar{G}$ to achieve satisfactory channel estimation performance and data transmission performance with affordable training overhead and circuit complexity.

\subsection{Multi-User Scenarios} 
The proposed channel estimation scheme can be easily generalized to multi-user cases by assuming all users transmit orthogonal pilot sequences such that channels from all users can be simultaneously estimated by the BS. 
This can be simply done by setting $K$ as the total number of antennas at all users. 
Alternatively, cascaded channels from different users can be successively estimated by assuming users transmitting pilots in a TDD manner. 
In this case, each user can transmit the same pilots and adopt the proposed channel estimation scheme to estimate the cascaded channel.   

\subsection{BD-RIS with Hybrid/Multi-Sector Modes}
\begin{figure}[t]
    \centering
    \includegraphics[width=0.48\textwidth]{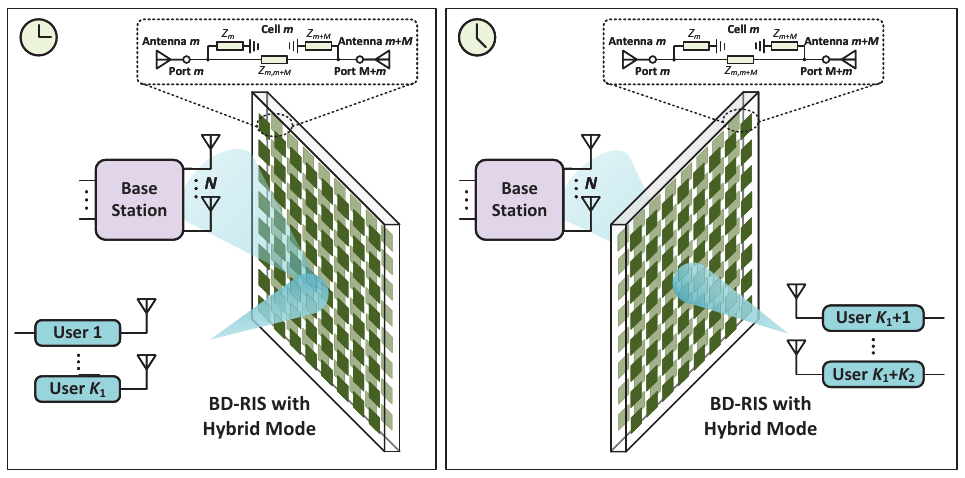}
    \caption{Illustration of channel estimation for BD-RIS with hybrid mode.}
    \label{fig:ce_hybrid}
\end{figure}

While the proposed channel estimation is based on BD-RIS with reflective mode, that is the user is located at the same side as BS, it is also readily generalized to BD-RIS with hybrid/multi-sector modes \cite{li2022,li2022beyond}. 
Assume that an $L$-sector BD-RIS with $M$ antennas in each sector, $L\in\mathcal{L} = \{1,\ldots,L\}, L\ge 2$, is adopted in a downlink MU-MISO system with an $N$-antenna BS and $K$ single-antenna users. Each sector covers $K_l$ users indexed by $\mathcal{K}_l = \{\sum_{i=1}^{l-1}K_i + 1, \cdots, \sum_{i=1}^lK_i\}$, $\sum_{l\in\mathcal{L}}K_l = K$.
In addition, the $L$-sector BD-RIS with group-connected architecture is, mathematically, characterized by matrices $\tilde{\mathbf{\Phi}}_l = \mathsf{blkdiag}(\tilde{\mathbf{\Phi}}_{l,1},\ldots,\tilde{\mathbf{\Phi}}_{l,G_1})\in\mathbb{C}^{M\times M}, \forall l\in\mathcal{L}$, satisfying the following constraint \cite{li2022beyond}
\begin{equation}\label{eq:multi_constraint}
    \sum_{l\in\mathcal{L}}\tilde{\mathbf{\Phi}}_{l,g}^H\tilde{\mathbf{\Phi}}_{l,g} = \mathbf{I}_{\bar{M}}, \forall g\in\mathcal{G}_1.
\end{equation}
Let $\mathbf{G} = [\mathbf{G}_1,\ldots,\mathbf{G}_{G_1}]\in\mathbb{C}^{N\times M}$ and $\mathbf{h}_k = [\mathbf{h}_{k,1}^T,\ldots,\mathbf{h}_{k,G_1}^T]^T\in\mathbb{C}^{M\times 1}$ denote the channel from BD-RIS to BS, and from user $k$, $\forall k\in\mathcal{K}$, to BD-RIS, respectively. 
Then the user-RIS-BS uplink channel can be expressed as 
\begin{equation}
    \label{eq:uplink_channel_hybmulti}
    \mathbf{h}_{\mathrm{u},k} = \sum_{g\in\mathcal{G}_1} \underbrace{\mathbf{h}_{k,g}^T\otimes\mathbf{G}_g}_{=\bar{\mathbf{Q}}_{k,g}\in\mathbb{C}^{N\times\bar{M}^2}} \mathsf{vec}(\tilde{\mathbf{\Phi}}_{l,g}), \forall k\in\mathcal{K}_l, \forall l\in\mathcal{L}.
\end{equation}
There are two strategies to estimate the above cascaded channels for users covered by different sectors. The first strategy is to successively estimate the channels for users covered by each sector. In this sense, the estimation process for each sector boils down to the case for reflective BD-RIS such that the proposed method can be directly adopted. 
The second strategy is to simultaneously estimate the channels from users covered by different sectors, while a new BD-RIS pattern design should be considered to adapt to the constraint (\ref{eq:multi_constraint}). Therefore, in this work, we focus on the first strategy and leave the second one as an interesting future work.
Applying the tile-based channel construction\footnote{For BD-RIS with hybrid/multi-sector mode, the \textit{group} is formed by the antenna ports from the same sector which are connected to each other. The \textit{tile} is formed by groups from the same sector.} proposed in Section \ref{sec:syst_mod}-B, we instead estimate the channels
\begin{equation}
    \label{eq:uplink_channel_stack_hybmulti}
    \mathbf{Q}_{k,i} = \sum_{j=1}^{\bar{G}}\bar{\mathbf{Q}}_{k,(i-1)\bar{G}+j}, \forall i\in\mathcal{G}_2, \forall k\in\mathcal{K}_l, \forall l\in\mathcal{L}.
\end{equation}
This can be done according to the following steps. 
\begin{enumerate}[S1:]
    \item In every $T_{1,l} = K_l\bar{M}^2G_2$ time durations, the $K_l$ cascaded channels 
    $\mathbf{Q}_{k,i}, \forall i\in\mathcal{G}_2, \forall k\in\mathcal{K}_l$,
    can be simultaneously estimated assuming 
    \begin{enumerate}[a)]
        \item that $\tilde{\mathbf{\Phi}}_{l'} = \mathbf{0}$, $\forall l'\ne l$, within the $T_{1,l}$ durations, 
        \item that $\tilde{\mathbf{\Phi}}_{l}$ and the pilots transmitted by $K_l$ users are designed using the proposed scheme. 
    \end{enumerate}
    This yields the channel estimates $\widehat{\mathbf{Q}}_{k,i}$, $\forall i\in\mathcal{G}_2$, $\forall k\in\mathcal{K}_l$.
    \item Repeat S1 for all $L$ sectors. 
\end{enumerate}
To facilitate understanding, we provide a simple example for how to estimate the cascaded channels for BD-RIS with hybrid mode ($L=2$) as illustrated in Fig. \ref{fig:ce_hybrid}. Based on the above strategy, we can obtain the cascaded channels for BD-RIS with hybrid/multi-sector modes with a total training overhead $T_1 = \sum_{l\in\mathcal{L}}T_{1,l} = \sum_{l\in\mathcal{L}}K_l\bar{M}^2G_2 = K\bar{M}^2G_2 = KM\bar{M}\bar{G}^{-1}$.

\section{Beamforming Design for BD-RIS}
\label{sec:beamforming}

In this section, we formulate two beamforming design problems for BD-RIS with reflective and hybrid/multi-sector modes and different scenarios, and propose efficient beamforming design methods relying on the cascaded channel estimates. The considered problems are different from that in previous works \cite{li2022,nerini2022optimal} since, in this work, only cascaded channel estimate is available, while the proposed algorithms \cite{li2022,nerini2022optimal} rely on separated BS-RIS and RIS-user channels. 

\subsection{Reflective BD-RIS Aided Point-to-Point MIMO}
We first consider a point-to-point MIMO system aided by a BD-RIS with reflective mode. 
Assume the $N$-antenna BS transmits the symbol vector $\mathbf{s}\in\mathbb{C}^{N_\mathrm{s}\times 1}$ to the $K$-antenna user, $\mathbb{E}\{\mathbf{s}\mathbf{s}^H\} = \mathbf{I}_{N_\mathrm{s}}$, where $N_\mathrm{s}$ denotes the number of data streams, $N_\mathrm{s} \le \min\{N,K\}$. 
Based on (\ref{eq:downlink_channel_stack}) and (\ref{eq:theta_stack}), the processed signal at the user is given by 
\begin{equation}
    \begin{aligned}
        \widehat{\mathbf{s}} 
        &= \mathbf{W}^H\mathbf{H}_\mathrm{d}\mathbf{P}\mathbf{s} + \mathbf{W}^H\mathbf{n}',
    \end{aligned}
\end{equation} 
where $\mathbf{W}\in\mathbb{C}^{K\times N_\mathrm{s}}$ and $\mathbf{P}\in\mathbb{C}^{N\times N_\mathrm{s}}$ denote the combiner at the user and the precoder at the BS, respectively, and $\mathbf{n}'\in\mathbb{C}^{K\times 1}$ with $\mathbf{n}'\sim\mathcal{CN}(\mathbf{0},\sigma'^2\mathbf{I}_K)$ denotes the noise. 

\subsubsection{Problem Formulation}
Based on the channel estimate $\widehat{\mathbf{q}} = \mathsf{vec}(\widehat{\mathbf{Q}})$, $\widehat{\mathbf{Q}} = [\widehat{\mathbf{Q}}_1,\cdots,\widehat{\mathbf{Q}}_{G_2}]$ with $\widehat{\mathbf{Q}}_i$ being the estimate of $\mathbf{Q}_i$, the joint transceiver and BD-RIS design problem to maximize the spectral efficiency is given by\footnote{Here, and in the following subsection, we involve the term $1 - \frac{T_1+T_2}{T}$ to account for both channel estimation efficiency and data transmission accuracy, while we ignore the effect of channel estimation error for simplicity and directly solve the optimization problem based on channel estimates. However, the effect of channel estimation error will be taken into account in the performance evaluations.} 
\begin{subequations}\label{eq:opt_prob}
    \begin{align}
    \non
    \max_{\bm{\theta},\mathbf{W},\mathbf{P}} ~~ &\Big(1 - \frac{T_1+T_2}{T}\Big)\log_2\Big(\Big|\mathbf{I}_{N_\mathrm{s}} + (\sigma'^2\mathbf{W}^H\mathbf{W})^{-1}\mathbf{W}^H\\
    \label{eq:obj_mimo}
    &~~\times \widehat{\mathbf{H}}_\mathrm{d}\mathbf{P}\mathbf{P}^H\widehat{\mathbf{H}}_\mathrm{d}^H\mathbf{W}\Big|\Big)\\
    \label{eq:ris_constraint}
    \mathrm{s.t.} ~~~&\mathbf{\Theta}_i^H\mathbf{\Theta}_i = \mathbf{I}_{\bar{M}}, \forall i\in\mathcal{G}_2,\\
    \label{eq:power_constraint}
    &\|\mathbf{P}\|_F^2 \le P_\mathrm{d},
    \end{align}
\end{subequations}
where $P_\mathrm{d}$ denotes the transmit power at the BS and $\widehat{\mathbf{H}}_\mathrm{d} = \mathsf{unvec}^T(\widehat{\mathbf{Q}}\bm{\theta})$, $\bm{\theta} = [\mathsf{vec}^T(\mathbf{\Theta}_1^T),\dots,\mathsf{vec}^T(\mathbf{\Theta}_{G_2}^T)]^T$. 
Problem (\ref{eq:opt_prob}) is not easy to solve due to the non-convex constraint (\ref{eq:ris_constraint}) and the coupling of the BD-RIS constraint, the precoder, and the combiner. 
To simplify the joint design, the objective function is decoupled into two separate optimizations. Specifically, we first design the BD-RIS to maximize the channel strength. Then, having the effective channel, we determine the precoder and the combiner to further maximize the spectral efficiency.  

\subsubsection{Solution to Problem (\ref{eq:opt_prob})}

With determined $\bm{\theta}$, it is well-known that the optimal unconstrained precoder and combiner for point-to-point MIMO are obtained from the unitary right and left singular vectors of the effective channel $\widehat{\mathbf{H}}_\mathrm{d}$.
Specifically, for the case of $N=K=N_\mathrm{s}$ and equal power allocation for all streams at the BS, we have $\mathbf{P}\mathbf{P}^H \propto \mathbf{I}_{N}$ and $\mathbf{W}\mathbf{W}^H \propto\mathbf{I}_{K}$. 
This motivates us to have the approximations $\mathbf{W}(\mathbf{W}^H\mathbf{W})^{-1}\mathbf{W}^H \approx \mathbf{I}_K$ and $\mathbf{P}\mathbf{P}^H \approx \zeta^2\mathbf{I}_N$, where $\zeta^2$ denotes the normalization factor. 
Based on the above discussions, the objective function in (\ref{eq:obj_mimo}) has an upper bound as 
\begin{equation}
    \begin{aligned}
        &\log_2\Big(\Big|\mathbf{I}_{N_\mathrm{s}} + \sigma'^{-2}\zeta^2\widehat{\mathbf{H}}_\mathrm{d}\widehat{\mathbf{H}}_\mathrm{d}^H\Big|\Big)\\
        &\overset{\text{(a)}}{=}\log_2\Big(\prod_{n=1}^{N_\mathrm{s}}(1+\lambda_n)\Big) = \sum_{n=1}^{N_\mathrm{s}}\log_2(1+\lambda_n)\\
        &\overset{\text{(b)}}{\le} \frac{1}{\ln2}\sum_{n=1}^{N_\mathrm{s}}\lambda_n = \frac{\zeta^2}{\ln2\sigma'^2}\mathsf{tr}(\widehat{\mathbf{H}}_\mathrm{d}\widehat{\mathbf{H}}_\mathrm{d}^H),
    \end{aligned}\label{eq:obj_mimo_simp}
\end{equation}
where (a) holds by defining $\lambda_1,\ldots,\lambda_{N_\mathrm{s}}$ as the eigenvalues of the matrix $\sigma'^{-2}\zeta^2\widehat{\mathbf{H}}_\mathrm{d}\widehat{\mathbf{H}}_\mathrm{d}^H$, (b) holds since $\lambda_n\ge 0$, $\forall n$.
In addition, the objective function in (\ref{eq:obj_mimo}) has a lower bound as
\begin{equation}
    \begin{aligned}
    &\log_2\Big(\prod_{n=1}^{N_\mathrm{s}}(1+\lambda_n)\Big) \\
    &= \log_2\Big(1 + \sum_{n=1}^{N_\mathrm{s}}\lambda_n + \sum_{n=1}^{N_\mathrm{s}}\sum_{n'=1}^{N_\mathrm{s}}\lambda_n\lambda_{n'} + \ldots\Big)\\
    &\ge\log_2\Big(1 + \sum_{n=1}^{N_\mathrm{s}}\lambda_n\Big) = \log_2\Big(1 + \frac{\zeta^2}{\sigma'^2}\mathsf{tr}(\widehat{\mathbf{H}}_\mathrm{d}\widehat{\mathbf{H}}_\mathrm{d}^H)\Big).
    \end{aligned}
\end{equation}
The above derivations indicate that maximizing the sum of eigenvalues helps in increasing both the lower and upper bounds of the rate. This motivates us to develop a two-stage design, where we first design the BD-RIS to maximize the channel strength, i.e., sum of eigenvalues, and then design the precoder and combiner to manage the inter-stream interference once $\bm{\theta}$ is determined. As such, we formulate the following BD-RIS design problem
\begin{subequations}\label{eq:solve_phi}
    \begin{align}
    \bm{\theta}^\star &= \mathop{\arg~\max}_{\mathbf{\Theta}_i^H\mathbf{\Theta}_i = \mathbf{I}_{\bar{M}}, \forall i} ~\mathsf{tr}(\widehat{\mathbf{H}}_\mathrm{d}\widehat{\mathbf{H}}_\mathrm{d}^H)\\
    \label{eq:obj}
    &\overset{\text{(a)}}{=} \mathop{\arg~\max}_{\mathbf{\Theta}_i^H\mathbf{\Theta}_i = \mathbf{I}_{\bar{M}}, \forall i} ~ \sum_{i'\in\mathcal{G}_2}\sum_{i\in\mathcal{G}_2}\bm{\theta}_i^T\mathbf{V}_{i,i'}\bm{\theta}_{i'}^*,
    \end{align}
\end{subequations}
where (a) follows with $\bm{\theta}_i = \mathsf{vec}(\mathbf{\Theta}_i^T)$ and $\mathbf{V}_{i,i'} = [\widehat{\mathbf{Q}}^T\widehat{\mathbf{Q}}^*]_{\mathcal{I}_{i},\mathcal{I}_{i'}}$, $\mathcal{I}_i = (i-1)\bar{M}^2+1:i\bar{M}^2$, $\forall i,i'\in\mathcal{L}$.
Note that $\bm{\theta}_i$ and $\bm{\theta}_{i'}$, $\forall i\ne i'$, $i,i'\in\mathcal{G}_2$, have separate constraints, while they are coupled in the objective function (\ref{eq:obj}), which makes it difficult to determine $\bm{\theta}_i$, $\forall i\in\mathcal{G}_2$. 
To simplify the design, we split the objective with respect to each $\bm{\theta}_i$ out from (\ref{eq:obj}) with fixed others. 
This results in the 
sub-problem for $\bm{\theta}_i$ as
\begin{subequations}\label{eq:opt_design_phi_l}
    \begin{align}
        \max_{\bm{\theta}_i} ~ &\bm{\theta}_i^T\mathbf{V}_{i,i}\bm{\theta}_i^* + 2\Re\{\bm{\theta}_i^T\bm{\chi}_i\}\\
        \label{eq:uni_constraint}
        \text{s.t.} ~~&\mathsf{unvec}^*(\bm{\theta}_i)\mathsf{unvec}^T(\bm{\theta}_i) = \mathbf{I}_{\bar{M}},
    \end{align}
\end{subequations}
where $\bm{\chi}_i=\sum\nolimits_{i'\ne i}\mathbf{V}_{i,i'}\bm{\theta}_{i'}^*\in\mathbb{C}^{\bar{M}^2\times 1}$, $\forall i\in\mathcal{G}_2$. Constraint (\ref{eq:uni_constraint}) constructs a $\bar{M}^2$-dimensional complex Stiefel manifold such that problem (\ref{eq:opt_design_phi_l}) can be formulated as an unconstrained optimization on the Stiefel manifold. Then, the problem can be effectively solved by the conjugate-gradient methods on the manifold space \cite{PAAbsil}.

After the BD-RIS has been designed, we obtain the effective channel as 
$\mathbf{H}_\mathrm{d}^\star = \mathsf{unvec}^T(\widehat{\mathbf{Q}}\bm{\theta}^\star)$. To obtain the precoder and combiner, we adopt the singular value decomposition (SVD) of the effective channel $\mathbf{H}_\mathrm{d}^\star$ as $\mathbf{H}_\mathrm{d}^\star = \mathbf{U}_1\mathbf{\Sigma}\mathbf{U}_2^H$, where $\mathbf{U}_1\in\mathbb{C}^{K\times K}$ and $\mathbf{U}_2\in\mathbb{C}^{N\times N}$ are unitary matrices, and $\mathbf{\Sigma}\in\mathbb{C}^{K\times N}$ is a rectangular diagonal matrix containing eigenvalues of $\mathbf{H}_\mathrm{d}^\star$ on the diagonal. Accordingly, we have the precoder and combiner as 
\begin{equation}
    \label{eq:precoder_combiner}
    \mathbf{P}^\star = \sqrt{P_\mathrm{d}}\frac{[\mathbf{U}_1]_{:,1:N_\mathrm{s}}}{\|[\mathbf{U}_1]_{:,1:N_\mathrm{s}}\|_F}, ~~\mathbf{W}^\star = [\mathbf{U}_2]_{:,1:N_\mathrm{s}}.
\end{equation}
For clarity, the above beamforming design algorithm is summarized in the following steps:
\begin{enumerate}[S1:]
    \item Initialize $\bm{\theta}$. 
    \item With fixed other terms, update $\bm{\theta}_i$ by solving problem (\ref{eq:opt_design_phi_l}).
    \item Repeat S2 until the convergence of $\bm{\theta}$ is guaranteed. 
    \item Obtain $\mathbf{P}^\star$ and $\mathbf{W}^\star$ by (\ref{eq:precoder_combiner}). 
\end{enumerate}

\subsection{Hybrid/Multi-Sector BD-RIS Aided MU-MISO}
In this subsection, we consider a MU-MISO system aided by a BD-RIS with hybrid/multi-sector mode, as illustrated in Section \ref{sec:CE}-C. Assume the $N$-antenna BS transmits the symbols $\tilde{\mathbf{s}} = [s_1,\ldots,s_K]\in\mathbb{C}^{K\times 1}$ to $K$ single-antenna users, $\mathbb{E}\{\tilde{\mathbf{s}}\tilde{\mathbf{s}}^H\} = \mathbf{I}_K$.
Based on (\ref{eq:uplink_channel_hybmulti}) and (\ref{eq:uplink_channel_stack_hybmulti}), the received signal at each user covered by sector $l$, $\forall l\in\mathcal{L}$, is given by 
\begin{equation}
    \begin{aligned}
        y_k =& \sum_{i\in\mathcal{G}_2}\mathsf{vec}^T(\mathbf{\Theta}_{l,i}^T)\mathbf{Q}_{k,i}^T\mathbf{p}_ks_k\\
        &+ \sum_{k'\ne k}\sum_{i\in\mathcal{G}_2}\mathsf{vec}^T(\mathbf{\Theta}_{l,i}^T)\mathbf{Q}_{k,i}^T\mathbf{p}_{k'}s_{k'} + n_k, \forall k\in\mathcal{K}_l,
    \end{aligned}
\end{equation}
where $\mathbf{\Theta}_{l,i}\in\mathbb{C}^{\bar{M}\times\bar{M}}$, $\forall l\in\mathcal{L}$, satisfies $\sum_{l\in\mathcal{L}}\mathbf{\Theta}_{l,i}^H\mathbf{\Theta}_{l,i} = \mathbf{I}_{\bar{M}}$, $\forall i\in\mathcal{G}_2$, $\mathbf{p}_k\in\mathbb{C}^{N\times 1}$ denotes the precoder for user $k$, and $n_k\in\mathbb{C}$ with $n_k\sim\mathcal{CN}(0,\sigma'^2)$ denotes the noise.   

\subsubsection{Problem Formulation} 
Based on the channel estimates $\widehat{\mathbf{Q}}_{k,i}$, $\forall k\in\mathcal{K}$, $\forall i\in\mathcal{G}_2$, the joint transmitter and BD-RIS design problem to maximize the spectral efficiency is
\begin{subequations}
    \label{eq:opt_phi_multi}
    \begin{align}
        \label{eq:obj_multi}
        \max_{\tilde{\bm{\theta}}_l,\forall l, \mathbf{p}_k,\forall k}~~&\Big(1-\frac{T_1+T_2}{T}\Big)\sum_{l\in\mathcal{L}}\sum_{k\in\mathcal{K}_l}\log_2(1 + \gamma_k)\\
        \text{s.t.} ~~~~~&\sum_{l\in\mathcal{L}}\mathbf{\Theta}_{l,i}^H\mathbf{\Theta}_{l,i} = \mathbf{I}_{\bar{M}}, \forall i\in\mathcal{G}_2.\\
        &\sum_{k\in\mathcal{K}}\|\mathbf{p}_k\|_2^2 \le P_\mathrm{d},
    \end{align}
\end{subequations}
where $\gamma_k = \frac{|\tilde{\bm{\theta}}_l^T\tilde{\mathbf{Q}}_{k}^T\mathbf{p}_k|^2}{\sum_{k'\ne k}|\tilde{\bm{\theta}}_l^T\tilde{\mathbf{Q}}_{k}^T\mathbf{p}_{k'}|^2+\sigma'^2}$ denotes the signal-to-interference-plus-noise ratio (SINR) with the fresh notations $\tilde{\bm{\theta}}_l = [\mathsf{vec}^T(\mathbf{\Theta}_{l,1}^T),\ldots,\mathsf{vec}^T(\mathbf{\Theta}_{l,G_2}^T)]^T$ and $\tilde{\mathbf{Q}}_k = [\widehat{\mathbf{Q}}_{k,1},\ldots,\widehat{\mathbf{Q}}_{k,G_2}]$, $\forall k\in\mathcal{K}_l$, $\forall l\in\mathcal{L}$.
Although problem (\ref{eq:opt_phi_multi}) is a challenging optimization with complicated ratio terms in the $\log(\cdot)$ function and non-convex constraints of BD-RIS, we show that the algorithm proposed in \cite{li2022} can be easily adapted to solving (\ref{eq:opt_phi_multi}) with slight modifications. 

\subsubsection{Solution to Problem (\ref{eq:opt_phi_multi})}
By introducing auxiliary variables $\iota_k$ and $\tau_k$, $\forall k\in\mathcal{K}_l$, $\forall l\in\mathcal{L}$ to each ratio term in (\ref{eq:obj_multi}) based on the fractional programming \cite{shen2018fractional}, we can transform (\ref{eq:obj_multi}) into a four-block optimization with a more decent objective function as $f(\{\tilde{\bm{\theta}}_l\}_{\forall l}, \{\mathbf{p}_k\}_{\forall k}, \{\iota_k\}_{\forall k}, \{\tau_k\}_{\forall k}) = \sum_{l\in\mathcal{L}}\sum_{k\in\mathcal{K}_l}(\log_2(1+\iota_k) - \iota_k + 2\sqrt{1+\iota_k}\Re\{\tau_k^*\tilde{\bm{\theta}}_l^T\tilde{\mathbf{Q}}_{k}^T\mathbf{p}_k\} - |\tau_k|^2(\sum_{k'\in\mathcal{K}}|\tilde{\bm{\theta}}_l^T\tilde{\mathbf{Q}}_{k}^T\mathbf{p}_{k'}|^2 + \sigma'^2))$. 
The four blocks are iteratively updated in sequential until convergence. 
Under the iterative design framework, the solutions to the blocks $\{\iota_k\}_{\forall k}$ and $\{\tau_k\}_{\forall k}$ can be easily obtained by checking the first-order conditions of $f(\{\tilde{\bm{\theta}}_l\}_{\forall l}, \{\mathbf{p}_k\}_{\forall k}, \{\iota_k\}_{\forall k}, \{\tau_k\}_{\forall k})$. Meanwhile, the solution to block $\{\mathbf{p}_k\}_{\forall k}$ can be obtained based on the Karush-Kuhn-Tucker (KKT) conditions. The detailed derivations can be done following \cite{shen2018fractional,li2022}. 

When blocks $\{\iota_k\}_{\forall k}$, $\{\tau_k\}_{\forall k}$, and $\{\mathbf{p}_k\}_{\forall k}$ are determined, the sub-problem for designing $\{\tilde{\bm{\theta}}_l\}_{\forall l}$ can be formulated as 
\begin{equation}
    \label{eq:solving_phi_multi}
    \begin{aligned}
        \{\tilde{\bm{\theta}}_l^\star\}_{\forall l} &= \mathop{\arg~\max}_{\sum\limits_{l\in\mathcal{L}}\mathbf{\Theta}_{l,i}^H\mathbf{\Theta}_{l,i} = \mathbf{I}_{\bar{M}}, \forall i} \sum_{l\in\mathcal{L}}\Big(2\Re\{\tilde{\bm{\theta}}_l^T\mathbf{v}_l\} - \tilde{\bm{\theta}}_l^T\mathbf{V}_l\tilde{\bm{\theta}}_l^*\Big)\\
        &= \mathop{\arg~\max}_{\sum\limits_{l\in\mathcal{L}}\mathbf{\Theta}_{l,i}^H\mathbf{\Theta}_{l,i} = \mathbf{I}_{\bar{M}}, \forall i} \sum_{l\in\mathcal{L}}\Big(2\sum_{i\in\mathcal{G}_2}\Re\{\bm{\theta}_{l,i}^T\mathbf{v}_{l,i}\}\\
        &~~~~~~ - \sum_{i\in\mathcal{G}_2}\sum_{i'\in\mathcal{G}_2}\bm{\theta}_{l,i}^T\mathbf{V}_{l,i,i'}\bm{\theta}_{l,i'}^*\Big),
    \end{aligned}
\end{equation}
where we define
$\mathbf{v}_l = \sum_{k\in\mathcal{K}_l}\sqrt{1+\iota_k}\tau_k^*\tilde{\mathbf{Q}}_k^T\mathbf{p}_k$,  $\mathbf{V}_l = \sum_{k\in\mathcal{K}_l}|\tau_k|^2\tilde{\mathbf{Q}}_k^T\sum_{k'\in\mathcal{K}}\mathbf{p}_{k'}\mathbf{p}_{k'}^H\tilde{\mathbf{Q}}_k^*$, $\forall l\in\mathcal{L}$, $\bm{\theta}_{l,i} = \mathsf{vec}(\mathbf{\Theta}_{l,i}^T)$, $\mathbf{v}_{l,i} = [\mathbf{v}_l]_{\mathcal{I}_i}$, and $\mathbf{V}_{l,i,i'} = [\mathbf{V}_l]_{\mathcal{I}_i,\mathcal{I}_{i'}}$, $\forall i,i'\in\mathcal{G}_2$.
Analogous to the derivation in Section \ref{sec:beamforming}-A, we propose to split the objective with respect to each $\{\bm{\theta}_{l,i}\}_{\forall l}$ from the objective in (\ref{eq:solving_phi_multi}) with fixed others. 
Defining $\ddot{\bm{\chi}}_{i} = [\bm{\chi}_{1,i}^T,\ldots,\bm{\chi}_{L,i}^T]^T$ with $\bm{\chi}_{l,i} = \mathbf{v}_{l,i} - \sum\nolimits_{i'\ne i}\mathbf{V}_{l,i,i'}\bm{\theta}_{l,i'}^*$, and $\ddot{\mathbf{V}}_i = \mathsf{blkdiag}(\mathbf{V}_{1,i,i},\ldots,\mathbf{V}_{L,i,i})$, 
we can construct the sub-problem for $\ddot{\bm{\theta}}_i = \mathsf{vec}([\mathbf{\Theta}_{1,i}^T,\ldots,\mathbf{\Theta}_{L,i}^T])$ as
\begin{subequations}
    \label{eq:sub_phi_i}
    \begin{align} 
        \max_{\ddot{\bm{\theta}}_i} ~&2\Re\{\ddot{\bm{\theta}}_{i}^T\ddot{\bm{\chi}}_i\}-\ddot{\bm{\theta}}_{i}^T\ddot{\mathbf{V}}_{i}\ddot{\bm{\theta}}_{i}^*\\
        \label{eq:constraint_ris_multi}
        \text{s.t.} ~~&\mathsf{unvec}^*(\ddot{\bm{\theta}}_i)\mathsf{unvec}^T(\ddot{\bm{\theta}}_i) = \mathbf{I}_{\bar{M}},
    \end{align}
\end{subequations}
where constraint (\ref{eq:constraint_ris_multi}) constructs a $\bar{M}^2L$-dimensional complex Stiefel manifold. Therefore, problem (\ref{eq:sub_phi_i}) can be solved by the manifold based methods \cite{PAAbsil}.

For clarity, the above beamforming design algorithm is summarized in the following steps:
\begin{enumerate}[S1:]
    \item Initialize $\tilde{\bm{\theta}}_l$, $\forall l\in\mathcal{L}$, and $\mathbf{p}_k$, $\forall k\in\mathcal{K}$.
    \item With fixed other terms, update $\{\iota_k\}_{\forall k}$ by solving problem $\{\iota_k^\star\}_{\forall k} = \arg\max~f(\{\tilde{\bm{\theta}}_l\}_{\forall l}, \{\mathbf{p}_k\}_{\forall k}, \{\iota_k\}_{\forall k}, \{\tau_k\}_{\forall k})$.
    \item With fixed other terms, update $\{\tau_k\}_{\forall k}$ by solving problem $\{\tau_k^\star\}_{\forall k} = \arg\max~f(\{\tilde{\bm{\theta}}_l\}_{\forall l}, \{\mathbf{p}_k\}_{\forall k}, \{\iota_k\}_{\forall k}, \{\tau_k\}_{\forall k})$.
    \item With fixed other terms, update $\{\mathbf{p}_k\}_{\forall k}$ by solving the constrained convex optimization problem $\{\mathbf{p}_k^\star\}_{\forall k} = \mathop{\arg\max}\limits_{\sum_{k\in\mathcal{K}}\|\mathbf{p}_k\|_2^2\le P_\mathrm{d}}f(\{\tilde{\bm{\theta}}_l\}_{\forall l}, \{\mathbf{p}_k\}_{\forall k}, \{\iota_k\}_{\forall k}, \{\tau_k\}_{\forall k})$.
    \item With fixed other terms, iteratively update $\ddot{\bm{\theta}}_i$, $\forall i\in\mathcal{G}_2$, by solving problem (\ref{eq:sub_phi_i}) until the convergence of $\{\ddot{\bm{\theta}}_i\}_{\forall i}$.
    \item Repeat S2-S5 until the convergence of the objective function $f(\{\tilde{\bm{\theta}}_l\}_{\forall l}, \{\mathbf{p}_k\}_{\forall k}, \{\iota_k\}_{\forall k}, \{\tau_k\}_{\forall k})$ is guaranteed.
\end{enumerate}

\section{Performance Evaluation}
\label{sec:simulation}

In this section, we perform simulation results to verify the effectiveness of the proposed channel estimation scheme and evaluate the performance of the proposed beamforming design.

\subsection{Simulation Setup}
The simulation setup is illustrated as follows.  
Both BS-RIS and RIS-user channels are assumed to have Rician fading with Rician factor $\kappa = 0$ dB accounting for the small-scale fading and distance-dependent path loss model in \cite{li2022beyond} accounting for the large-scale fading. The performance for channel estimation with Rayleigh fading channels can be found in \cite{li2023channel}.
Specifically, we assume each BD-RIS antenna has an idealized radiation pattern \cite{li2022beyond}, such that the path loss model is expressed as\footnote{For the case of BD-RIS with reflective mode, we calculate the path loss by setting the number of sectors as $L=2$, which indicates that BD-RIS with reflective mode covers half of the space.} $\xi = \varrho(1-\cos\frac{\pi}{L})^2$, where $\varrho = 4^3\pi^4d_1^{\epsilon_1}d_2^{\epsilon_2}\lambda^{-4}G_\mathrm{t}^{-1}G_\mathrm{r}^{-1}$ is calculated by the BS-RIS distance $d_1 = 30$ m and RIS-user distance $d_2 = 10$ m, path loss exponents $\epsilon_1 = 2.5$ and $\epsilon_2 = 2.5$, wavelength $\lambda = \frac{c}{f}$ with $f=2.4$ GHz, antenna gains at the BS side $G_\mathrm{t} = 1$ and the user side $G_\mathrm{r} = 1$. 
More details of the analysis for the BD-RIS aided path loss model can be found in \cite{li2022beyond}.
The noise powers for uplink channel estimation and downlink transmission are set as $\sigma^2=\sigma'^2=-100$ dBm.
In the following simulations, we set the group size $\bar{M}$ of the group-connected architecture for BD-RIS as $\bar{M}\in\{1,2,4\}$ with an affordable circuit complexity. More comprehensive studies of different BD-RIS architectures and their circuit complexities can be found in \cite{nerini2023beyond}.

\subsection{Performance for Channel Estimation}

\begin{figure}
    \centering
    \includegraphics[width=0.45\textwidth]{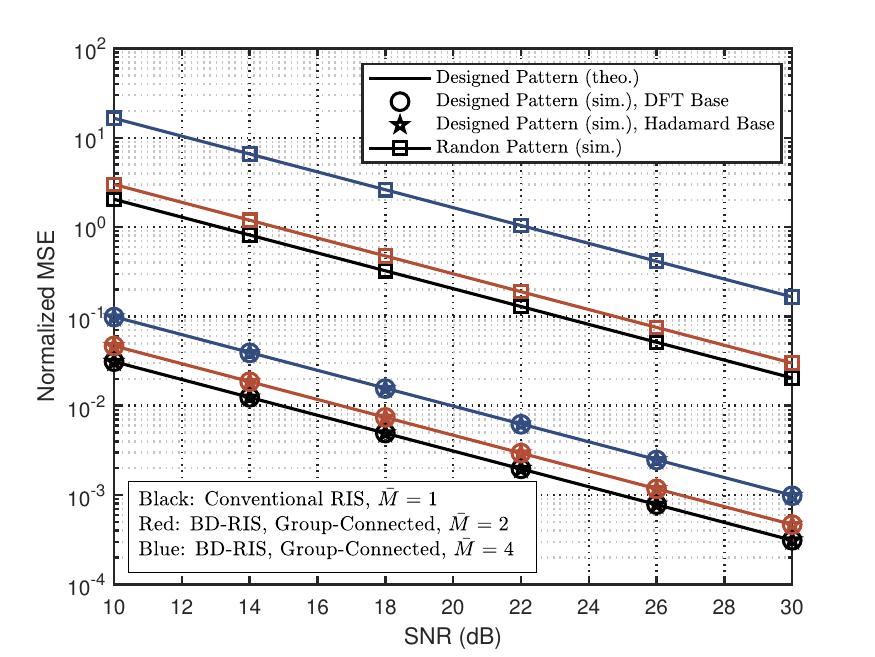}
    \caption{Normalized MSE versus SNR ($M = 32$, $\bar{G} = 4$, $N=K=2$).}
    \label{fig:MSE_P}
\end{figure}

We evaluate the channel estimation performance by plotting the normalized MSE versus signal-to-noise ratio (SNR) with different group size of group-connected BD-RIS and different numbers of BS/user antennas\footnote{Here we only plot the case of BD-RIS with reflective mode since, as explained in Section \ref{sec:CE}-E, the same scheme can be used in BD-RIS with hybrid/multi-sector modes with the same closed-form MSE performance. The effect of channel estimation error will be studied in the performance evaluation of beamforming design.}.
Specifically, the normalized MSE is calculated by 
$\bar{\mathrm{e}}_{\widehat{\mathbf{q}}} = \mathbb{E}\{\frac{\|\widehat{\mathbf{q}}-\mathbf{q}\|_2^2}{\|\mathbf{q}\|_2^2}\}$.
For fair comparison, the SNR is defined as the ratio between the total power received at each antenna from each transmit antenna through each RIS element for the whole uplink training and the noise power, i.e., $\bar{\gamma} = \frac{P_\mathrm{u}T_1}{\xi\sigma^2}$.
From Fig. \ref{fig:MSE_P}, we have the following observations\footnote{In Fig. \ref{fig:MSE_P}, the benchmark scheme, referred to as ``Random Pattern'', is generated by replacing S2-S6 of the proposed channel estimation method in Section \ref{sec:CE}-B with $\mathbf{\Phi}_\mathrm{rand}$. Specifically, $\mathbf{\Phi}_\mathrm{rand}$ is constructed with $[\mathbf{\Phi}_\mathrm{rand}]_{i,(l-1)\bar{M}^2+1:l\bar{M}^2} = \mathsf{vec}^T(\mathbf{U}_{i,l})$, where $\mathbf{U}_{i,l}\in\mathbb{C}^{\bar{M}\times\bar{M}}$, $\forall i,l$ are randomly generated unitary matrices.}. 
\textit{First}, the matrix $\widehat{\mathbf{\Phi}}$ with DFT base or Hadamard base to generate $\mathbf{Z}_1$, $\mathbf{Z}_2$, and $\mathbf{X}$ achieves exactly the same performance, which verifies Theorem 1. 
\textit{Second}, the matrix $\widehat{\mathbf{\Phi}}$ with DFT/Hadamard bases achieves exactly the same performance as the theoretical minimum, which verifies Lemma 2.  
\textit{Third}, the normalized MSE performance increases with the group size $\bar{M}$ of group-connected BD-RIS, which verifies Lemma 1.
This indicates that, with the same power budget, BD-RIS leads to higher channel estimation error than conventional RIS.  
The intuition behind is that the same amount of power is split to estimate more channel coefficients, which leads to increasing channel estimation error.
\textit{Fourth}, the normalized MSE decreases with SNR, which indicates that larger received SNR leads to smaller channel estimation error. 
\textit{Fifth}, the matrix $\widehat{\mathbf{\Phi}}$ designed by the proposed scheme achieves better MSE performance than that constructed by random unitary matrices, which demonstrates the effectiveness of the proposed design for channel estimation.

\begin{figure}
    \centering
    \includegraphics[width=0.45\textwidth]{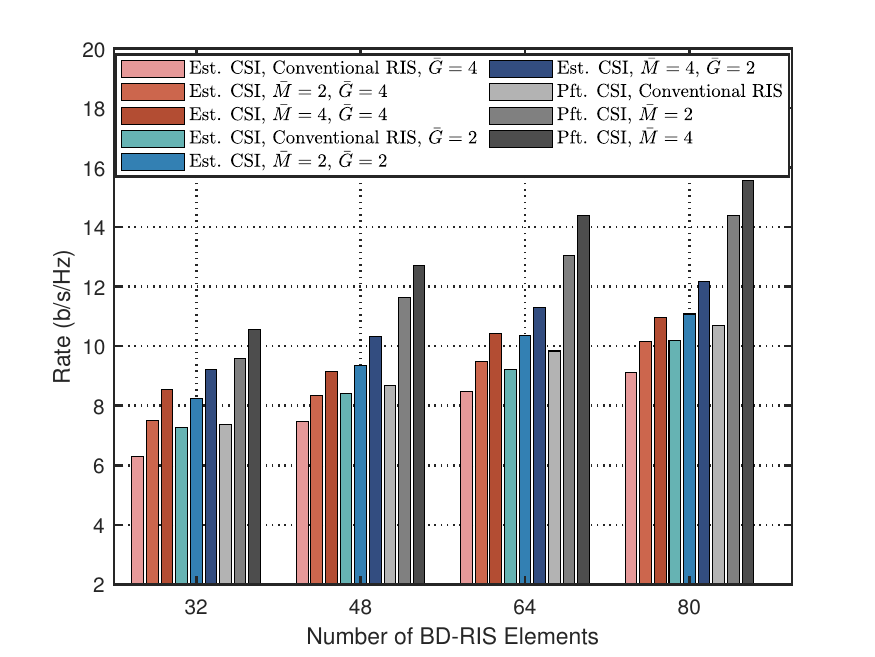}
    \caption{Rate versus the number of BD-RIS elements $M$($N = K = N_\mathrm{s} = 2$, $P_\mathrm{u} = 250$ mW, $P_\mathrm{d} = KP_\mathrm{u}$).}
    \label{fig:SE_M}
\end{figure}

\subsection{Performance for Beamforming Design}

\subsubsection{Reflective BD-RIS Aided Point-to-Point MIMO}
We start by evaluating the performance of the proposed beamforming design algorithm from Section \ref{sec:beamforming}-A in Fig. \ref{fig:SE_M}. 
For comparison, we simulate the performance achieved by perfect CSI $\mathbf{Q}$ as an upper bound. 
We also ignore the term $1 - \frac{T_1+T_2}{T}$ in the objective function (\ref{eq:obj_mimo}), that is, to evaluate the rate performance for the ease of analyzing the effect of beamforming design. 
From Fig. \ref{fig:SE_M}, we have the following observations. 
\textit{First}, the rate performance achieved based on both the perfect CSI and the estimated CSI increases with increasing group size $\bar{M}$ of the group-connected architecture, that is, BD-RIS outperforms conventional RIS, which aligns with the results in \cite{shen2021,li2022,nerini2022optimal}.
\textit{Second}, with the fixed group size $\bar{M}$, increasing the tile size $\bar{G}$ during the channel estimation induces the rate performance degradation due to the reduction of CSI, which further limits the BD-RIS design. 
\textit{Third}, the rate performance gap between the perfect CSI case and the estimated CSI case increases with increasing number of BD-RIS elements. This comes from the fact that more BD-RIS elements provide more flexibility in beamforming design, while the rate performance is more sensitive to CSI conditions.

\begin{figure}
    \centering
    \includegraphics[width=0.45\textwidth]{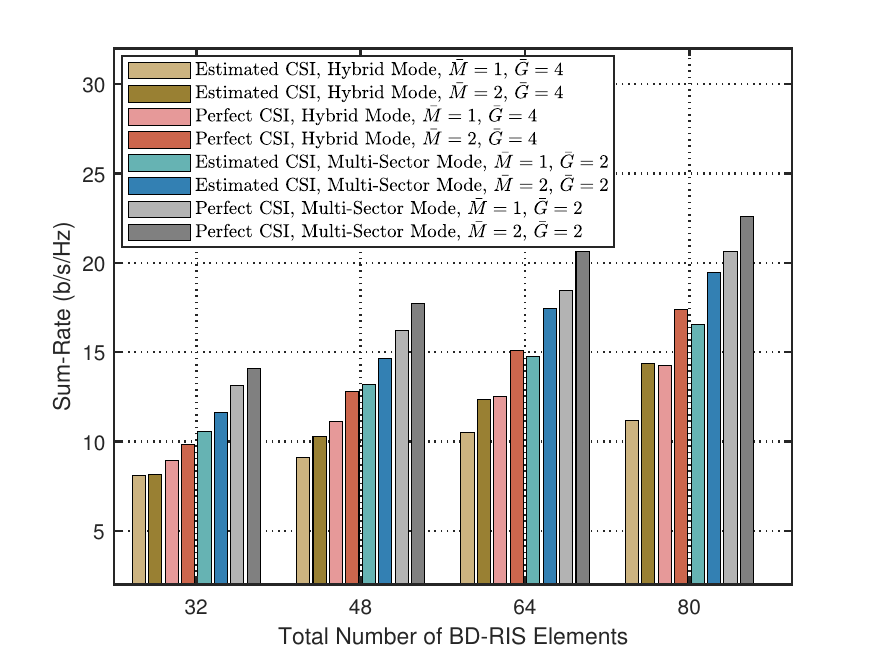}
    \caption{Sum-rate versus the total number of BD-RIS elements $ML$ ($N = K = 4$,  $P_\mathrm{u} = 250$ mW, $P_\mathrm{d} = KP_\mathrm{u}$).}
    \label{fig:SR_M}
\end{figure}

\subsubsection{Hybrid/Multi-Sector BD-RIS Aided MU-MISO}
We next evaluate the sum-rate performance of the proposed algorithm from Section \ref{sec:beamforming}-B in Fig. \ref{fig:SR_M}, where the term $1-\frac{T_1+T_2}{T}$ in the objective function (\ref{eq:obj_multi}) is also ignored to obtain the sum-rate. 
In addition, in the following simulations, we fix the total number of elements $ML$ for BD-RIS with hybrid mode ($L=2$) and multi-sector mode ($L=4$). For BD-RIS with hybrid and multi-sector modes, every $K/L$ users are located within the coverage of each sector. 
From Fig. \ref{fig:SR_M}, we have the following observations.
\textit{First}, for both perfect and estimated CSI cases, BD-RIS with multi-sector mode outperforms BD-RIS with hybrid mode thanks to the use of antennas with narrower beamwidth in the multi-sector mode, which aligns with the results in \cite{li2022beyond}.
\textit{Second}, the sum-rate performance achieved by BD-RIS with hybrid/multi-sector modes increases with $\bar{M}$ due to the increasing beam manipulation flexibility, which aligns with the results in \cite{li2023reconfigurable}.

\subsection{Overhead and Transmission Performance Trade-Off}

To get more insights on the trade-off between the training overhead for channel estimation and the performance for data transmission, we take into account the term $1-\frac{T_1 + T_2}{T}$ in the following simulations to evaluate the spectral efficiency.
For simplicity, we ignore the symbol duration for processing and feedback, i.e., $T_2 = 0$, since the channel estimation overhead $T_1$ and the symbol duration for data transmission $T_3$ are generally much larger than $T_2$. 
In addition, we set the symbol duration for one transmission frame $T$ as $T\in\{600,1000,2000\}$. In practice, the value of $T$ depends on multiple factors, such as the carrier frequency of transmit signals, phase differences in the multipath propagation, and mobility \cite{bjornson2017massive}.

\begin{figure}
    \centering
    \subfigure[Conventional RIS ($\bar{M} = 1$)]{
    \includegraphics[width=0.45\textwidth]{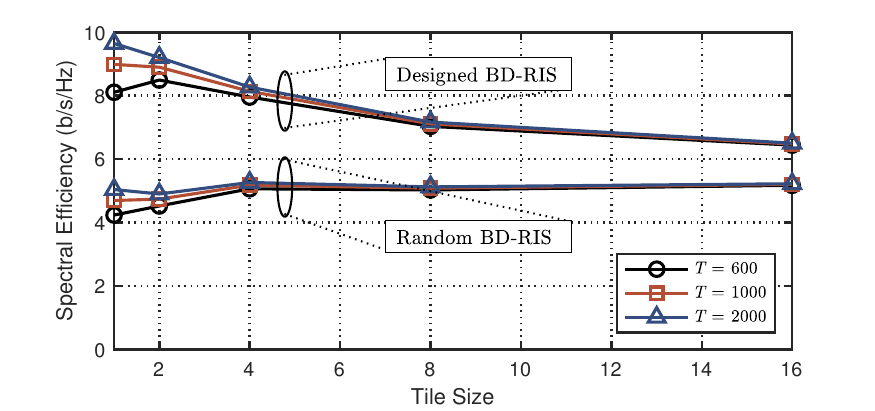}}
    \subfigure[BD-RIS ($\bar{M} = 2$)]{
    \includegraphics[width=0.45\textwidth]{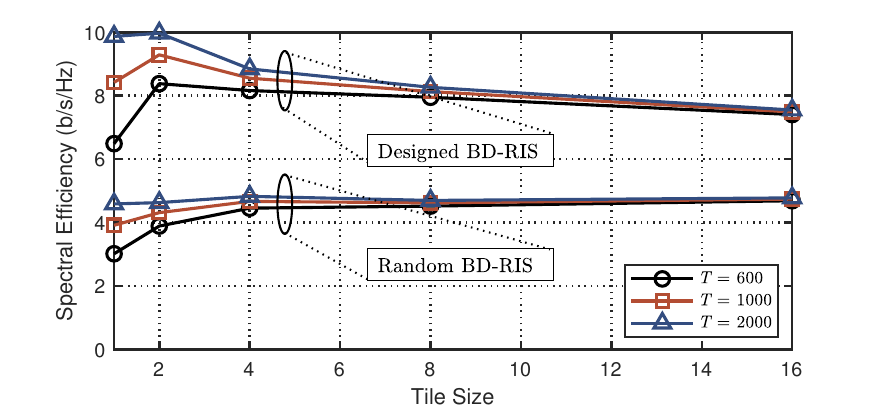}}
    \subfigure[BD-RIS ($\bar{M} = 4$)]{
    \includegraphics[width=0.45\textwidth]{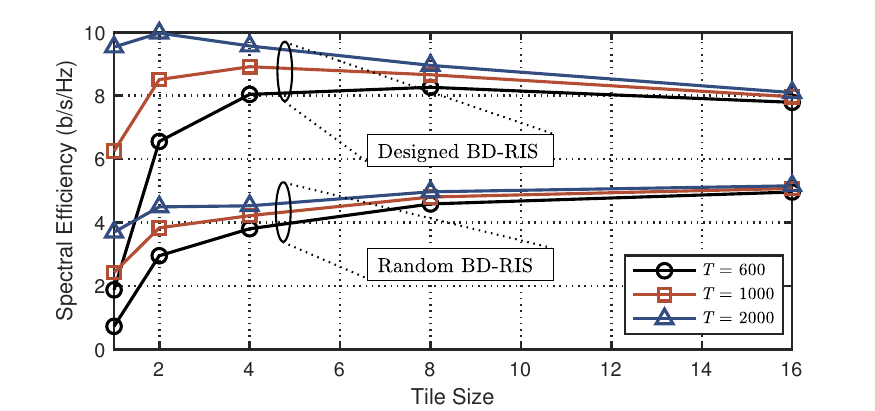}}
    \caption{Spectral efficiency versus tile size $\bar{G}$ ($M = 64$, $N = K = N_\mathrm{s} = 2$, $P_\mathrm{u} = 250$ mW, $P_\mathrm{d} = KP_\mathrm{u}$).}
    \label{fig:SE_barG}
\end{figure}

\begin{figure}
    \centering
    \subfigure[Single-Connected ($\bar{M} = 1$)]{
    \includegraphics[width=0.45\textwidth]{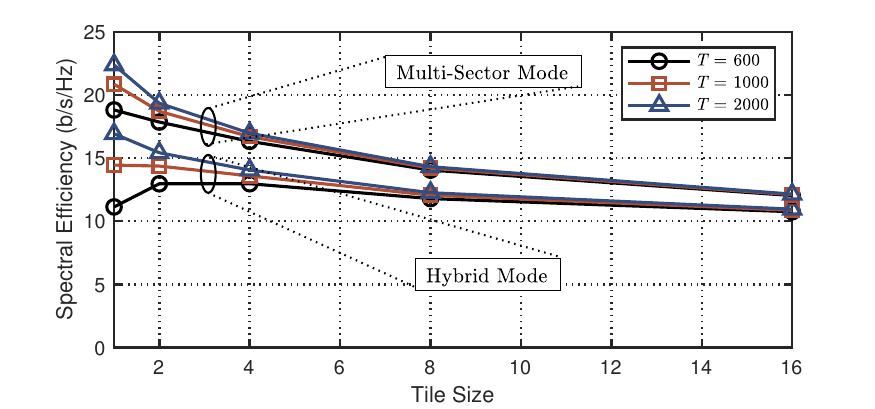}}
    \subfigure[Group-Connected ($\bar{M} = 2$)]{
    \includegraphics[width=0.45\textwidth]{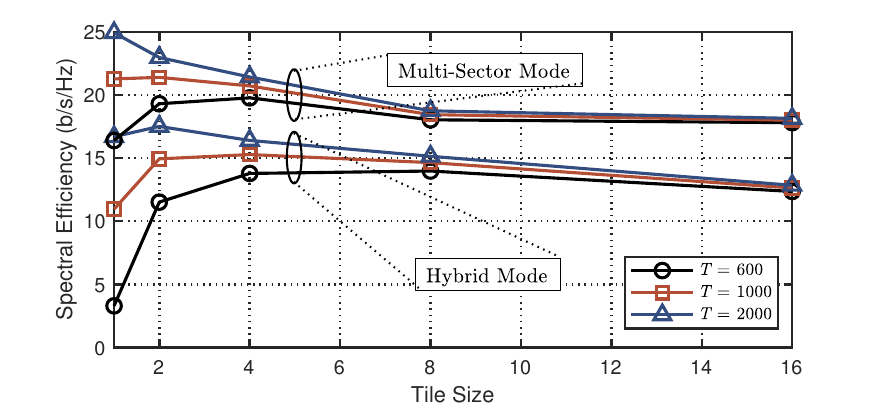}}
    \caption{Spectral efficiency versus tile size $\bar{G}$ ($ML = 128$, $N = K = 4$, $P_\mathrm{u} = 250$ mW, $P_\mathrm{d} = KP_\mathrm{u}$).}
    \label{fig:SR_barG}
\end{figure}

\subsubsection{Reflective BD-RIS Aided Point-to-Point MIMO}
We first plot the spectral efficiency defined in (\ref{eq:obj_mimo}) versus the tile size $\bar{G}$ for BD-RIS with different architectures in Fig. \ref{fig:SE_barG}, from which we have the following observations\footnote{In Fig. \ref{fig:SE_barG}, the benchmark scheme, referred to as ``Random BD-RIS'', is realized by estimating the channel based on the proposed method, randomly generating a block-diagonal unitary matrix $\mathbf{\Theta}$, and performing the SVD based on the effective channel for downlink transmission.}. 
\textit{First}, for a relatively small symbol duration, e.g., $T\in\{600, 1000\}$, the spectral efficiency achieved by the proposed design in general first increases and then decreases with the increasing tile size $\bar{G}$. This is because a larger tile size leads to a smaller training overhead $T_1$ but a reduced beam manipulation flexibility of BD-RIS. 
For example, when $T = 600$, the best spectral efficiency is achieved at $\bar{G} = 2$ for group-connected BD-RIS with $\bar{M} = 2$, and at $\bar{G} = 8$ with $\bar{M} = 4$.
In this case, the rate performance improvement by more flexible BD-RIS design is overwhelmed by the large training overhead. 
\textit{Second}, for a relatively large symbol duration, e.g. $T = 2000$, the spectral efficiency achieved by the proposed design decreases with the increasing tile size. 
This is because the large $T$ weakens the impact of the long training overhead at the small tile size during channel estimation, while the rate performance keeps decreasing with increasing tile size. 
\textit{Third}, with fixed BD-RIS architectures and training overhead, a larger symbol duration leads to better spectral efficiency performance. 
These observations highlight the importance of adjusting the time slot allocation for uplink and downlink to ensure satisfactory overall performance. 
\textit{Fourth}, for a relatively small tile size, e.g. $\bar{G}\in\{1,2\}$, the spectral efficiency achieved by BD-RIS with a larger group size performs worse due to the significant training overhead consumption in channel estimation, which overwhelms the rate performance enhancement provided by more flexible BD-RIS architectures.
\textit{Fifth}, for a relatively large tile size, e.g. $\bar{G}\in\{8,16\}$, the spectral efficiency achieved by BD-RIS outperforms the conventional RIS thanks to the additional beam manipulation flexibility supported by the group-connected architecture.
Therefore, the tile size $\bar{G}$ and $\bar{M}$ should be carefully chosen to balance the training overhead for channel estimation, the rate performance, and the circuit complexity for BD-RIS implementation.

\subsubsection{Hybrid/Multi-Sector BD-RIS Aided MU-MISO}
In Fig. \ref{fig:SR_barG}, we plot the spectral efficiency defined in (\ref{eq:obj_multi}) as a function of $\bar{G}$ for BD-RIS with hybrid/multi-sector modes. 
From Fig. \ref{fig:SR_barG} we observe that, in addition to the findings obtained from Fig. \ref{fig:SE_barG}, BD-RIS with multi-sector mode is more robust to the symbol durations $T$ and the tile size. This is because the significant channel gain enhancement provided by high-gain antennas compensates for the effects of the training overhead and the beamforming flexibility.

\section{Conclusion}
\label{sec:conclusion}

In this paper, we propose a novel channel estimation scheme for BD-RIS aided multi-antenna systems. Specifically, the BD-RIS has different (single/group/fully-connected) architectures and (reflective/hybrid/multi-sector) modes, which, mathematically, generate unique constraints and complicate the channel estimation. To tackle this difficulty, we propose an efficient pilot sequence and BD-RIS design to estimate the cascaded BD-RIS channel, which achieves the minimum MSE of the LS estimator. 
The analysis and extensions of the proposed channel estimation scheme are also illustrated. 

With the cascaded channel estimates, we consider two BD-RIS scenarios, that is reflective BD-RIS aided point-to-point MIMO, and hybrid/multi-sector BD-RIS aided MU-MISO, and propose efficient algorithms to design the transceiver and BD-RIS with different architectures and modes.

Simulation results demonstrate the advantages of the proposed design. 
Specifically, the effectiveness of the proposed channel estimation scheme and beamforming algorithms are verified by comparing with the design for perfect CSI cases. 
In addition, the trade-off between the training overhead for channel estimation and the performance for data transmission is analyzed with different parameter settings. 

Beyond this work, future research avenues include faster and more efficient channel estimation methods for BD-RIS by exploring the circuit topologies of different architectures to reduce the dimensionality and for active RISs \cite{zhi2022active}.

\begin{appendices}
    \section{Proof of Lemma 1}
    The objective (\ref{eq:obj_MSE}) has the following lower bound 
    \begin{subequations}\label{eq:HM_AM}
        \begin{align}
        &\mathsf{tr}((\widehat{\mathbf{\Phi}}^H\widehat{\mathbf{\Phi}})^{-1})
        = \sum_{i=1}^{NK\bar{M}^2G_2}[(\widehat{\mathbf{\Phi}}^H\widehat{\mathbf{\Phi}})^{-1}]_{i,i} \\ &\overset{\text{(a)}}{\ge}\sum_{i=1}^{NK\bar{M}^2G_2}[\widehat{\mathbf{\Phi}}^H\widehat{\mathbf{\Phi}}]_{i,i}^{-1}\\
        &\overset{\text{(b)}}{\ge} \frac{N^2K^2G_2^2\bar{M}^4}{\sum_{i=1}^{NK\bar{M}^2G_2}[\widehat{\mathbf{\Phi}}^H\widehat{\mathbf{\Phi}}]_{i,i}}
        = \frac{N^2K^2G_2^2\bar{M}^4}{\mathsf{tr}(\widehat{\mathbf{\Phi}}^H\widehat{\mathbf{\Phi}})}\\
        &= \frac{N^2K^2G_2^2\bar{M}^4}{\mathsf{tr}(\sum_{t=1}^{K\bar{M}^2G_2}(\bar{\bm{\phi}}_t^*\otimes\bar{\mathbf{x}}_t^*\otimes\mathbf{I}_N)(\bar{\bm{\phi}}_t^T\otimes\bar{\mathbf{x}}_t^T\otimes\mathbf{I}_N))}\\
        &\overset{\text{(c)}}{=} \frac{N^2K^2G_2^2\bar{M}^4}{\mathsf{tr}(\sum_{t=1}^{K\bar{M}^2G_2}(\bar{\bm{\phi}}_t^*\bar{\bm{\phi}}_t^T)\otimes(\bar{\mathbf{x}}_t^*\bar{\mathbf{x}}_t^T)\otimes\mathbf{I}_N)}\\
        &\overset{\text{(d)}}{=} \frac{NKG_2^2\bar{M}^4}{\sum_{t=1}^{K\bar{M}^2G_2}\mathsf{tr}(\bar{\bm{\phi}}_t^*\bar{\bm{\phi}}_t^T)}\\
        &\overset{\text{(e)}}{=} N\bar{M},
        \end{align}
    \end{subequations}
    where the equality of (a) can be attained when $\widehat{\mathbf{\Phi}}^H\widehat{\mathbf{\Phi}}$ is a diagonal matrix \cite{kay1993fundamentals};
    (b) holds due to the relationship between the harmonic mean and the arithmetic mean with equality achieved when $\widehat{\mathbf{\Phi}}^H\widehat{\mathbf{\Phi}}$ has identical diagonal entries; (c) holds by the property of the Kronecker product;
    (d) holds due to the property of the trace operation and the assumption that $|[\mathbf{x}_t]_k| = 1$, $\forall k$, yielding $\mathsf{tr}(\mathbf{x}_t^*\mathbf{x}_t^T) = K$, $\forall t$; 
    (e) holds due to the constraint (\ref{eq:unitary_constraint}) which yields $\mathsf{tr}(\bar{\bm{\phi}}_t^*\bar{\bm{\phi}}_t^T) = G_2\bar{M}$, $\forall t$. Based on the above derivations, we have that $\min \mathsf{tr}((\widehat{\mathbf{\Phi}}^H\widehat{\mathbf{\Phi}})^{-1})=N\bar{M}$ with the condition $\widehat{\mathbf{\Phi}}^H\widehat{\mathbf{\Phi}} = \widehat{\mathbf{\Phi}}\widehat{\mathbf{\Phi}}^H=K\bar{M}G_2\mathbf{I}_{NK\bar{M}^2G_2}$, which completes the proof.

    \section{Proof of Lemma 2}
    With $\mathbf{A}$ satisfying (\ref{eq:fx_cons1}), (\ref{eq:fx_cons2}) and $\breve{\mathbf{\Phi}}$ satisfying (\ref{eq:fc1_cons1}), (\ref{eq:fc1_cons2}), we have $\mathbf{\Phi} = \mathbf{A}\otimes\breve{\mathbf{\Phi}}$ such that
    \begin{equation}
        \begin{aligned}
            \mathbf{\Phi}^H\mathbf{\Phi} &= (\mathbf{A}^H\otimes\breve{\mathbf{\Phi}}^H)(\mathbf{A}\otimes\breve{\mathbf{\Phi}})
            = (\mathbf{A}^H\mathbf{A})\otimes(\breve{\mathbf{\Phi}}^H\breve{\mathbf{\Phi}})\\
            &= (G_2\mathbf{I}_{G_2})\otimes(\bar{M}\mathbf{I}_{\bar{M}^2})
            = G_2\bar{M}\mathbf{I}_{G_2\bar{M}^2},
        \end{aligned}
    \end{equation} 
    which aligns with (\ref{eq:fc_cons1}).

    In addition, we define the $i$-th sub-block rows of the $s$-th row of $\mathbf{\Phi}$ by $\bm{\phi}_{s,i}$, $\forall s\in\mathcal{S}$, $\forall i\in\mathcal{G}_2$, i.e., 
    \begin{equation}
        \bm{\phi}_{s,i}=[\mathbf{\Phi}]_{s,(i-1)\bar{M}^2:i\bar{M}^2}=[\bm{\phi}_{s}^T]_{(i-1)\bar{M}^2:i\bar{M}^2}.
    \end{equation} 
    Meanwhile, from (\ref{eq:block_training}) we have 
    \begin{equation}
        \bm{\phi}_{s,i}=[\bar{\bm{\phi}}_t^T]_{(i-1)\bar{M}^2:i\bar{M}^2}=\mathsf{vec}^T(\mathbf{\Phi}_{t,i}),
    \end{equation} 
    where $t=(s-1)K+k$, $\forall k\in\mathcal{K}$.
    Furthermore, from Lemma 2 we have $\bm{\phi}_{s,i}=[\mathbf{A}]_{i',i}[\breve{\mathbf{\Phi}}]_{m,:}$, where $s = (i'-1)\bar{M}^2+m$, $\forall m\in\bar{\bar{\mathcal{M}}}$.
    Hence, (\ref{eq:unitary_constraint}) is satisfied, i.e.,
    \begin{equation}
        \begin{aligned}
        \mathbf{\Phi}_{t,i}^H\mathbf{\Phi}_{t,i} 
        &= |[\mathbf{A}]_{i',i}|^2 \breve{\mathbf{\Phi}}_m^H\breve{\mathbf{\Phi}}_m
        =\mathbf{I}_{\bar{M}},
        \end{aligned}
    \end{equation} 
    where $\breve{\mathbf{\Phi}}_m = \mathsf{unvec}([\breve{\mathbf{\Phi}}^T]_{:,m})$.
    The proof is completed.

    \section{Proof of Theorem 1}
    We start by proving $\breve{\bm{\phi}}_{m,n}$, $\forall m,n\in\bar{\mathcal{M}}$ satisfies (\ref{eq:fc1_cons2}). To this end, we define $\breve{\bm{\phi}}_{m,n,i} = [\breve{\bm{\phi}}_{m,n}]_{(i-1)\bar{M}+1:i\bar{M}}$, $\forall i\in\bar{\mathcal{M}}$. We have 
    \begin{equation}
        \breve{\bm{\phi}}_{m,n,i} = [\mathbf{Z}_2]_{m,i}[\mathbf{Z}_1^T]_{\mathsf{mod}(i-n,\bar{M})+1,:}.
    \end{equation} 
    Then we calculate
    \begin{equation}
        \begin{aligned}
            \breve{\bm{\phi}}_{m,n,i}^*\breve{\bm{\phi}}_{m,n,i'}^T
            =&\underbrace{[\mathbf{Z}_1^H]_{\mathsf{mod}(i-n,\bar{M})+1,:}[\mathbf{Z}_1]_{:,\mathsf{mod}(i'-n,\bar{M})+1}}_{=z_{1,i.i',n}}\\
            &\times\underbrace{[\mathbf{Z}_2]_{m,i}^*[\mathbf{Z}_2]_{m,i'}}_{=z_{2,i,i',m}},
        \end{aligned}
    \end{equation}
    yielding the following two conditions:
    \begin{enumerate}[1)]
        \item when $i=i'$, we have $z_{1,i,i,n} = \alpha_1$ and $z_{2,i,i,m} = \frac{\alpha_2}{\bar{M}}$ such that $\breve{\bm{\phi}}_{m,n,i}^*\breve{\bm{\phi}}_{m,n,i'}^T=1$;
        \item when $i\ne i'$, we have $z_{1,i,i',n} = 0$ such that $\breve{\bm{\phi}}_{m,n,i}^*\times \breve{\bm{\phi}}_{m,n,i'}^T = 0$. 
    \end{enumerate}
    Therefore, (\ref{eq:fc1_cons2}) is guaranteed.

    We next prove $\breve{\mathbf{\Phi}}$ constructed by $\breve{\bm{\phi}}_{m,n}$ satisfies (\ref{eq:fc1_cons1}). To this end, we calculate 
    \begin{equation}
        \begin{aligned}
            \breve{\bm{\phi}}_{m,n}^*\breve{\bm{\phi}}_{m',n'}^T 
            =& \sum_{i\in\bar{\mathcal{M}}}\breve{\bm{\phi}}_{m,n,i}^*\breve{\bm{\phi}}_{m',n',i}^T\\
            =& \sum_{i\in\bar{\mathcal{M}}}\underbrace{[\mathbf{Z}_1^H]_{\mathsf{mod}(i-n,\bar{M})+1,:}[\mathbf{Z}_1]_{:,\mathsf{mod}(i-n',\bar{M})+1}}_{=z_{1',n,n',i}}\\
            &~~~~~\times\underbrace{[\mathbf{Z}_{2}]_{m,i}^*[\mathbf{Z}_{2}]_{m',i}}_{=z_{2',m,m',i}},
        \end{aligned}
    \end{equation}
    yielding the following three conditions:
    \begin{enumerate}[1)]
        \item when $m=m'$ and $n=n'$, we have $z_{1',n,n,i} = \alpha_1$ and $z_{2',m,m,i} = \frac{\alpha_2}{\bar{M}}$ such that $\breve{\bm{\phi}}_{m,n}^*\breve{\bm{\phi}}_{m,n}^T = \bar{M}$; 
        \item when $n\ne n'$, we have $z_{1',n,n',i} =0$ such that $\breve{\bm{\phi}}_{m,n}^*\times \breve{\bm{\phi}}_{m',n'}^T = 0$;
        \item when $m\ne m'$ and $n=n'$, we have $z_{1',n,n,i} = \alpha_1$ such that $\breve{\bm{\phi}}_{m,n}^*\breve{\bm{\phi}}_{m',n'}^T = \alpha_1\sum_{i\in\bar{\mathcal{M}}}z_{2',m,m',i} = \alpha_1[\mathbf{Z}_2^*]_{m,:}[\mathbf{Z}_2^T]_{:,m'} = 0$. 
    \end{enumerate}
    Therefore, (\ref{eq:fc1_cons1}) is guaranteed. The proof is completed.

\end{appendices}

\bibliographystyle{IEEEtran}
\bibliography{refs}
	
\end{document}